\documentclass{article}

\usepackage{arxiv}

\usepackage[utf8]{inputenc} 
\usepackage[T1]{fontenc}    
\usepackage{hyperref}       
\usepackage{url}            
\usepackage{booktabs}       
\usepackage{amsfonts}       
\usepackage{nicefrac}       
\usepackage{microtype}      
\usepackage{lipsum}		
\usepackage{graphicx}
\usepackage{natbib}
\usepackage{doi}

\usepackage{amsmath}
\usepackage{bm}
\usepackage{amssymb}
\usepackage{hhline}
\usepackage{subcaption}
\usepackage{comment}

\usepackage{titlesec}
\usepackage{color}
\usepackage[table]{xcolor}
\usepackage{tabularx}
\usepackage{rotating}
\usepackage{graphicx}

\DeclareMathOperator{\Tr}{Tr}

\usepackage{algorithm}
\usepackage{algpseudocode}
\definecolor{customblue}{rgb}{0,0,1}

\usepackage{hhline}
\usepackage{graphicx}
\graphicspath{{Images/}}
\usepackage{eso-pic}

\usepackage[overload]{empheq}
\definecolor{deeppink}{rgb}{1.0, 0.08, 0.58}
\usepackage{comment} 
\usepackage{fancyhdr} 
\usepackage{lipsum} 
\usepackage{tcolorbox} 
\usepackage{colortbl}
\setcitestyle{authoryear,round}

\usepackage{float} 

\title{Uncover mortality patterns and hospital effects in COVID-19 heart failure patients: a novel Multilevel logistic cluster-weighted modeling approach}

\author{ Luca Caldera  \\
	MOX, Department of Mathematics, Politecnico di Milano, Piazza Leonardo da Vinci 32, Milano (IT) \\
	\texttt{luca.caldera@polimi.it} \\
	\AND
	Chiara Masci  \\
	MOX, Department of Mathematics, Politecnico di Milano, Piazza Leonardo da Vinci 32, Milano (IT) \\
	\texttt{chiara.masci@polimi.it} \\
 \AND
	Andrea Cappozzo  \\
	Department of Economics, Management and Quantitative Methods, Università degli Studi di Milano, \\ Via Conservatorio 7, Milano (IT) \\
	\texttt{andrea.cappozzo@unimi.it} \\
 \AND
	Marco Forlani  \\
	 Regione Lombardia, Informatica SPA, Piazza Città di Lombardia 1, Milano (IT) \\
	\texttt{marco.forlani@ext.ariaspa.it} \\
 \AND
	Barbara Antonelli  \\
	Regione Lombardia, Divisione Servizi per il Welfare Regionale, Piazza Città di Lombardia 1, Milano (IT) \\
	\texttt{barbara.antonelli@ariaspa.it} \\
 \AND
 Olivia Leoni  \\
	U.O. Osservatorio Epidemiologico, DG Welfare, Regione Lombardia, Piazza Città di Lombardia 1, Milano (IT) \\
	\texttt{olivia\_leoni@regione.lombardia.it} \\
 \AND
	Francesca Ieva  \\
	MOX, Department of Mathematics, Politecnico di Milano, Piazza Leonardo da Vinci 32, Milano (IT) \\
 Health Data Science Centre, Human Technopole, Viale Rita Levi‑Montalcini 1, Milano (IT) \\
	\texttt{francesca.ieva@polimi.it} \\
}

\renewcommand{\shorttitle}{Uncover mortality patterns and hospital effects in COVID-19 HF patients: a novel ML-CWM modeling approach}

\hypersetup{
pdftitle={A template for the arxiv style},
pdfsubject={q-bio.NC, q-bio.QM},
pdfauthor={David S.~Hippocampus, Elias D.~Striatum},
pdfkeywords={First keyword, Second keyword, More},
}

\begin{document}
\maketitle

\begin{abstract}
	Evaluating hospitals' performance and its relation to patients' characteristics is of utmost importance to ensure timely, effective, and optimal treatment. Such a matter is particularly relevant in areas and situations where the healthcare system must contend with an unexpected surge in hospitalizations, such as for heart failure patients in the Lombardy region of Italy during the COVID-19 pandemic. Motivated by this issue, the paper introduces a novel Multilevel Logistic Cluster-Weighted Model (ML-CWMd) for predicting 45-day mortality following hospitalization due to COVID-19. The methodology flexibly accommodates dependence patterns among continuous, categorical, and dichotomous variables; effectively accounting for hospital-specific effects in distinct patient subgroups showing different attributes. A tailored Expectation-Maximization algorithm is developed for parameter estimation, and extensive simulation studies are conducted to evaluate its performance against competing models. The novel approach is applied to administrative data from the Lombardy Region, aiming to profile heart failure patients hospitalized for COVID-19 and investigate the hospital-level impact on their overall mortality. A scenario analysis demonstrates the model's efficacy in managing multiple sources of heterogeneity, thereby yielding promising results in aiding healthcare providers and policy-makers in the identification of patient-specific treatment pathways.
\end{abstract}


\section{Introduction}
\label{s:intro}
Medical data often possess a hierarchical structure and encompass hidden clusters of patients. To evaluate the effectiveness of the health care system or assess the impact of a certain treatment or pathology on patient outcomes, classical techniques are often incapable of jointly taking into account the heterogeneity given by the hospital where the patient is treated and the intrinsic characteristics of the patient. This holds significance as failing to separate the hospital effect from the individual patient's factors limits the policy's practicability. Consequently, the necessity intensifies for the adoption of sophisticated analytical tools capable of comprehensively addressing these multifaceted factors. By embracing these tools, the medical field is better poised to extract richer insights, accounting for the intricate interplay of variables that collectively shape patient outcomes \citep{copss2012statistical, ammenwerth2003evaluation, Berta2016}. Motivated by the utilization of healthcare administrative data, this study focuses on profiling and categorizing heart-failure patients from Lombardy region, Italy, who have been hospitalized for COVID-19. The significance of studying such patients stems from the observed correlation between COVID-19 and heart failure, which during the pandemic revealed a notably high mortality rate. The coexistence of these conditions frequently hinders the recovery process for patients infected with the virus \citep{rey2020heart, bader2021heart}.
Moreover, the virus itself can also trigger heart failure in patients who have contracted it \citep{adeghate2021mechanisms}. Consequently, the objective of the present work is multifold: firstly, assess the combined impact of hospitals and individual patient characteristics to enhance patient profiling, especially for vulnerable populations like those afflicted by COVID-19, and to evaluate hospital structures accordingly. Secondly, we are dedicated to examining and analyzing how respiratory illnesses affect clinical results and general well-being of patients within this particular demographic. This understanding will be crucial in developing tailored interventions to enhance patient outcomes. The proposed approach originates from the general framework of mixture modeling, that handles observations nested within groups, specifically within the realm of cluster-weighted models \cite[CWM,][]{Gershenfeld1997a,Ingrassia2012, Ingrassia2014}. CWMs are employed to represent the joint distribution of a random vector including a response variable and a set of covariates in situations in which the data can be naturally divided into clusters. Stemming from the original formulation of CWM, numerous research directions have been explored, leading to extensions that encompass, but are not limited to, various response types \citep{Ingrassia2015,Punzo2016a}, high-dimensional settings \citep{Subedi2013, Subedi2015}, robust estimation \citep{Ingrassia2014, Garcia-Escudero2017, Perrone2024}, and multivariate regression \citep{Dang2017a,Tomarchio2021,Gallaugher2022}. Additionally, an active line of research has been devoted to capitalizing on the flexibility inherent in the CWM framework within the context of hierarchical data, addressing both continuous \citep{Berta2016} and dichotomous responses \citep{Berta2019}.

In details, leveraging on the multilevel logistic cluster-weighted model (ML-CWM) introduced in \cite{Berta2019}, we devise a novel ML-CWM accounting for various types of dependence, denoted with ML-CWMd hereafter. Specifically, the proposed model captures the dependence between observations belonging to the same cluster and group and expands the set of covariates to accommodate dichotomous dependent variables making use of the Ising model \citep{Ghosal2020,cheng2014sparse}. Modeling the interdependence of binary variables is essential in the medical field because it allows for the analysis of relationships between the presence or absence of various diseases, or among different risk factors observed in patients. These enhancements enable the model to encapsulate a broader spectrum of data dynamics and relationships, providing a group specific risk assessment and subsequently offering a more insightful understanding of the underlying processes at play. The primary focus of the ML-CWMd is the medical field, particularly in cases involving hierarchical data at a single group level. Throughout this paper, we refer to these groups as ``hospitals'' and the individual data points within each group as ``patients''. This terminology highlights the model's applicability in the medical domain, where hierarchical groupings often correspond to hospital structures. Note that the effectiveness of multilevel models in this context is also highlighted by their successful application in analyzing COVID-19 data, as demonstrated in recent studies \citep{Verbeeck2023, Berta2024}. 

The remainder of the paper is structured as follows. In Section~\ref{s:model}, we introduce the ML-CWMd model. In Section~\ref{s: EM-algo}, we outline the EM-algorithm necessary for model estimation. Section~\ref{sec: sim study expl}  encompasses a simulation study aimed at comprehending the efficacy of the proposed model. In Section~\ref{s: application}, we showcase the application of the model to Lombardy region healthcare administrative data, complete with a scenario analysis. Section~\ref{s:discuss} draws the conclusions. The proposed model has been implemented in \texttt{R} \citep{RCoreTeam}, and the developed routines are freely available and accessible at the following GitHub repository: \href{https://github.com/luca2245/ML-CWMd}{https://github.com/luca2245/ML-CWMd}. The repository also includes a Shiny application, enabling clinician users to access model estimates by uploading their own patients' data and automatically visualize results and predictions. This application can potentially serve as a valuable tool for integrating this modeling approach into clinical practice.

\section{Methodology}
\label{s:model}

Consider a dichotomous response variable Y and a set of covariates \textbf{X} = (\textbf{U},\textbf{V},\textbf{D}), where \textbf{U} represents a $p-$dimensional vector of continuous variables, \textbf{V} represents a $q-$dimensional vector of categorical variables and \textbf{D} represents an $h-$dimensional vector of dichotomous variables which may possess some degree of dependence. In a two-levels hierarchy, units $i$, for $i = 1, \dots, n_j$, are nested within groups $j$, for $j = 1, \dots, J$.  Consider \textbf{X} and Y defined in a finite space $\bm{\Omega}$ with values in $\mathbb{R}^{(p + q + h)}$ × $\{0,1\}$, which is assumed to be divided into $C$ clusters denoted as $\bm{\Omega}_1,\hdots, \bm{\Omega}_C$.
Given this setup, the joint probability across the clusters can be factorized as follows \citep{Berta2019}:
\begin{align}
\label{eq:joint density}
p((\mathbf{x},\mathbf{y})| \bm{\theta}) &= \sum_{c= 1}^{C}
p(\mathbf{y}|\mathbf{x},\bm{\xi}_{c})\phi(\mathbf{u}|
\bm{\mu}_{c},\bm{\Sigma}_c) \psi(\mathbf{v}|\bm{\lambda}_{c}) \zeta(\bm{d}| \bm{\Gamma}_c,\bm{\nu}_c)  w_{c} 
\end{align}

\noindent
where $\bm{\theta}$ is the vector containing all the model parameters and $w_c$ indicates the proportion of observations within cluster $c$. We model the continuous covariates $\mathbf{U}$ using a multivariate normal density $\phi(\cdot|\boldsymbol{\mu}_c,\boldsymbol{\Sigma}_c)$, with cluster-wise different mean vectors $\boldsymbol{\mu}_c$ and covariance matrices $\boldsymbol{\Sigma}_c$; the independent categorical covariates $\mathbf{V}$, each potentially possessing a varying number of categories, using $q$ independent multinomial distributions $\psi(\cdot|\bm{\lambda}_{c})$ with cluster-wise different parameter vectors $\boldsymbol{\lambda}_{c}$; the dichotomous dependent covariates $\mathbf{D}$ using the Ising model presented in \cite{Ghosal2020}, characterized by distribution $\zeta(\cdot| \bm{\Gamma}_c,\bm{\nu}_c)$, with cluster-wise different threshold vectors $\bm{\nu}_c$ and interaction matrices $\bm{\Gamma}_c$. It is assumed that $\mathbf{U}$, $\mathbf{V}$, and $\mathbf{D}$ exhibit local independence within the clusters. Within each cluster, $p(\mathbf{y}|\mathbf{x},\bm{\xi}_{c})$ represents a multilevel logistic regression model where $\bm{\xi}_c$ are the cluster-specific parameters for both fixed and random effects. Without loss of generality, we hereafter discuss the setting in which covariates in the conditional and marginal distributions of \eqref{eq:joint density} are identical, although in practical scenarios they might vary partially or entirely (see Section \ref{s: application}). This variation leads to what is referred to in the literature as mixtures of regression with concomitant variables \citep{Dayton1988}. In details, in each cluster the conditional distribution takes on the following form:
\begin{align}
\label{eq: logistic regr 0} 
p(\mathbf{y}|\mathbf{x},\bm{\xi}_{c}) &=  \big{[} \bm{\pi}_{j}  \big{]}^{\bm{y}_{j}} \big{[} 1 - \bm{\pi}_{j}  \big{]}^{1 - \bm{y}_{j}}
\end{align}
\begin{align}
\label{eq: logistic regr} 
\text{logit}(\bm{\pi}_{j}) = \textbf{F}_j \ \bm{\beta}_c +  \textbf{R}_j \ \bm{b}_{j,c}; \quad \quad \bm{\pi}_{j} = \frac{ \exp\{\textbf{F}_j \ \bm{\beta}_c +  \textbf{R}_j \ \bm{b}_{j,c}\} }{ 1 + \exp\{\textbf{F}_j \ \bm{\beta}_c +  \textbf{R}_j \ \bm{b}_{j,c}  \} }
\end{align}

\noindent
where $\text{logit}(\bm{\pi}_{j})$ represents the odds associated with group $j$, indicating the probability of an event occurring divided by the probability of not occurring. The coefficients $\bm{b}_{j,c} \sim \mathcal{N}(\bm{0},\textbf{D}_c)$ represent the random effects for group $j$ in cluster $c$. The vector $\bm{\beta}_c$ comprises the coefficients corresponding to fixed effects. Matrices $\textbf{F}_j$ and $\textbf{R}_j$ represent the design matrices associated with fixed effects and random effects in group $j$, respectively. We denote the design matrix encompassing all types of covariates as $\textbf{X}_j = (\textbf{F}_j \cup \textbf{R}_j)$. All specifications regarding the other terms outlined in Equation~\ref{eq:joint density} will be detailed in subsequent sections.

\subsection{Model for Categorical Covariates V}
\label{subsec: model V}
Consider $q$ categorical covariates indexed by $r = 1,\dots,q$. In each cluster, we model the categorical covariates \textbf{V} with $q$ independent multinomial distributions of parameter $\bm{\lambda}_{cr}$ ($c = 1,\dots,C$, $r = 1,\dots,q$), identifying the vector of probabilities for the $k_r$ categories of the $r-$th covariate. Here we are making the assumption that every categorical variable takes on the $k_r$ categories across clusters $c = 1,\dots,C$. In this case, we can represent each of these covariates using a binary vector $\bm{v}^r = (v^{r1}, \dots, v^{rk_r})$ such that \citep{Ingrassia2015}:
\begin{align}
\label{eq:binary}
v^{rs} = \begin{cases}
1, & \text{if } {v}^r = s, s \in \{1,\dots,k_r \}, \\
0, & \text{otherwise}.
\end{cases}
\end{align}
Therefore, the density $\psi$, given by the product of $q$ conditionally independent multinomial distributions, can be written as:
\begin{align}
\label{eq:multinom distribution}
\psi(\bm{v}|\bm{\lambda}_{c}) = \prod_{r = 1}^{q} \prod_{s = 1}^{k_r}  {\lambda_{crs}}^{v^{rs}} \quad c = 1,\dots,C.
\end{align}

\subsection{Model for Dichotomous Dependent Covariates D}
\label{subsec: model D}
Consider $h$ dichotomous variables indexed by $l = 1,\dots,h$ which may possess some degree of dependence and let $\bm{d}_{ij} = ({d}_{ij1},\dots,{d}_{ijh})$ be the $h-$dimensional binary random vector associated to observation $i$ and group $j$ within cluster c. The probability of observing this vector is specified by the Ising model with cluster-specific parameters $\bm{\Gamma}_c$ and $\bm{\nu}_c$ as follows:
\begin{align}
\label{eq:zeta density}
\zeta(\bm{d}_{ij}|\bm{\Gamma}_c,\bm{\nu}_c) =  \frac{1}{S(\bm{\Gamma}_c,\bm{\nu}_c)} \exp \bigg{(}\frac{1}{2}\bm{d}_{ij}^{T} \bm{\Gamma}_{c} \bm{d}_{ij} + \bm{d}_{ij}^{T} \bm{\nu}_c \bigg{)}\quad c = 1,...,C
\end{align}
\begin{flalign*}
& \bm{\nu}_c =  [\nu_1^c, \dots,\nu_h^c] \quad ; \quad \bm{\Gamma}_{c} =  \begin{bmatrix}
 0 & \gamma^c_{1,2} & \hdots & \gamma^c_{1,h}\\
\gamma^c_{2,1} & \gamma^c_{2,2} & \hdots & \gamma^c_{2,h} \\
\vdots & \vdots & \ddots & \vdots \\
 \gamma^c_{h,1} & \gamma^c_{h,2} & \dots & 0
\end{bmatrix} 
\end{flalign*}
$\bm{\Gamma}_{c} \in \mathbb{R}^{h\text{x}h}$ is a symmetric matrix and $S$ is the normalizing constant. 
The interaction parameters $\bm{\Gamma}$ depict the interrelationships between all pairs of binary variables. Meanwhile, the threshold parameters $\bm{\nu}$ highlight the inclination of a variable to lean towards one state or the other when all interaction parameters related to that particular variable are equal to zero. The Ising model has formulations in two domains (\{0,1\} and \{-1,1\}), each offering a different interpretation of interaction parameters. Within the \{0, 1\} domain, augmenting the interaction parameter between two variables (e.g., $\gamma_{12}$), results in an increased likelihood of the state (1, 1) when compared to the other feasible states: (0, 0), (0, 1), and (1, 0). Conversely, within the \{-1, 1\} domain, elevating the interaction parameter between the two variables gives rise to a higher probability of the states (1, 1) and (-1, -1), in contrast to the states (-1, 1) and (1, -1). Based on the assumed pattern of interdependence between the dichotomous variables, one can choose to utilize either of the two domains. See Web Appendix A for further information about the Ising model.

\subsection{Likelihood of the Model}
\label{subsec: model likelihood}
Consider a sample of $N = n_1 + \cdots + n_J$  observation pairs $ \{(\mathbf{x}_{ij},y_{ij})\}_{j = 1,\dots,J, \  i = 1, \dots,n_j}$ drawn from model \eqref{eq:joint density}. To express the likelihood, we introduce a latent variable $\bm{z}$ as
\begin{align}
\label{eq:latent variables}
z_{ijc} = \begin{cases}
1, & \text{if } (\mathbf{x}_{ij}, y_{ij}) \ \text{belongs to the $c-$th clusters} \\
0, & \text{otherwise}
\end{cases}
\end{align}
and we make the assumption of independence between observations belonging to different clusters and independence between observations within the same cluster but belonging to different groups (highest-level units). With a slight abuse of notation, we use the vector $\bm{z}_{jc}$ to indicate that only observations within group $j$ that are part of cluster $c$ contribute to the likelihood associated with cluster $c$ and we express the complete likelihood as
\begin{align}
\label{eq:likelihood 2}
L((\mathbf{x},\mathbf{y}, \mathbf{z}); \bm{\theta}) &=   \prod_{c= 1}^{C} \prod_{j = 1}^{J}\bigg{[}p(\mathbf{y}_{j}|\mathbf{x}_{j},\bm{\xi}_{c})\phi(\mathbf{u}_{j}|\bm{\mu}_{c},\bm{\Sigma}_c)\psi(\mathbf{v}_{j}|\bm{\lambda}_{c}) \zeta(\bm{d}| \bm{\Gamma}_c,\bm{\nu}_c) w_{c} \bigg{]}^{\mathbf{z}_{jc}}
\end{align}
and the complete log-likelihood as
\begin{align}
&l((\mathbf{x},\mathbf{y}, \mathbf{z}); \bm{\theta}) = \sum_{c= 1}^{C} \sum_{j = 1}^{J}   \mathbf{z}_{jc} \ \log\big{(}p(\mathbf{y}_{j}|\mathbf{x}_{j},\bm{\xi}_{c}) \phi(\mathbf{u}_{j}|\bm{\mu}_{c},\bm{\Sigma}_c) \psi(\mathbf{v}_{j}|\bm{\lambda}_{c}) \zeta(\bm{d}_j| \bm{\Gamma}_c,\bm{\nu}_c) w_{c}\big{)}. \label{eq::log-likelihood 1} 
\end{align}
Since we take into account the dependence between observations in the same cluster and group only in the regression component, we can factorise the log-likelihood as follows:
\begin{align}
l((\mathbf{x},\mathbf{y}, \mathbf{z}); \bm{\theta}) &= \sum_{c= 1}^{C} \sum_{j = 1}^{J}  \mathbf{z}_{jc} \log\big{(}p(\mathbf{y}_{j}|\mathbf{x}_{j},\bm{\xi}_{c})\big{)} +
\sum_{c= 1}^{C} \sum_{j = 1}^{J} \sum_{i = 1}^{n_j}  {z}_{ijc} \log\big{(}\phi(\mathbf{u}_{ij}|\bm{\mu}_{c},\bm{\Sigma}_c)\big{)} + \notag \\
 & + \sum_{c= 1}^{C} \sum_{j = 1}^{J} 
\sum_{i = 1}^{n_j} {z}_{ijc} \log\big{(}\psi(\mathbf{v}_{ij}|\bm{\lambda}_{c})\big{)} +\sum_{c= 1}^{C} \sum_{j = 1}^{J} \sum_{i = 1}^{n_j} {z}_{ijc} \log\big{(}\zeta(\bm{d}_{ij}| \bm{\Gamma}_c,\bm{\nu}_c)\big{)} + \notag \\
& \sum_{c= 1}^{C} \sum_{j = 1}^{J} \sum_{i = 1}^{n_j}   {z}_{ijc} \log\big{(}w_{c}\big{)}. \label{eq::log-likelihood 2} 
\end{align}
\\
In particular, for the regression component, we assume independence:
\vspace{1mm}
\begin{itemize}
         \item[{{$\bullet$}}] between observations originating from different clusters;

        \item[{{$\bullet$}}] between observations originating from same cluster but belonging to different groups.

\end{itemize}
The first assumption offers the advantage of simplifying the log-likelihood form, facilitating its maximization as elaborated in the next section. However, this assumption comes with a drawback, i.e., it precludes the model from capturing potential associations among observations belonging to the same group (e.g., hospital) but to different clusters. 
In the health context, while it might seem reasonable to assume that patients treated in the same hospital are interconnected, it is also worth noting that these patients, belonging to distinct clusters, exhibit diverse needs and behaviours. 
Conversely, the association between individuals belonging to the same hospital and to the same cluster is assumed to be adequately accounted for by the random effects incorporated within the regression component of the model.

\section{Model Estimation}
\label{s: EM-algo}

For the estimation of the model parameters, we devise a tailored Expectation-Maximization \citep[EM algorithm, ][]{Dempster1977, McLachlan2008}. In the following, we define the E and the M steps that are alternated until convergence. 

\subsection{E-step}
\label{subsec: E-step}
At the $(k+1)-$th iteration, the E-step involves computing the expected value of the log-likelihood $l((\mathbf{x},\mathbf{y}, \mathbf{z}); \bm{\theta})$ defined in Equation \eqref{eq::log-likelihood 2}, given the observed data and the previous estimate $\hat{\bm{\theta}}^{(k)}$, obtained from iteration $k$. Therefore, the E-step entails calculating the expectation of the random variables $Z_{ijc}$ associated to $z_{ijc}$. In particular, for $j = 1,\dots,J, i = 1,\dots , n_j, c = 1,\dots ,C$, we have:

\begin{align}
\mathbb{E}\big{[} z_{ijc}|(\mathbf{x},\mathbf{y}), \hat{\bm{\theta}}^{(k)} \big{]} = \mathbb{P}\big{[} z_{ijc} = 1|(\mathbf{x}_{ij},y_{ij}, ), \hat{\bm{\theta}}^{(k)} \big{]} =  \tau_{ijc}^{(k)}((\mathbf{x}_{ij},y_{ij}), \hat{\bm{\theta}}^{(k)}) = \notag \\
= \frac{p(y_{ij}|\mathbf{x}_{ij},\hat{\bm{\xi}}_{c}^{(k)})\phi(\mathbf{u}_{ij}|\hat{\bm{\mu}}_{c}^{(k)},\hat{\bm{\Sigma}}_c^{(k)}) \psi(\mathbf{v}_{ij}|\hat{\bm{\lambda}}_{c}^{(k)}) \zeta(\bm{d}_{ij}| \hat{\bm{\Gamma}}_c^{(k)},\hat{\bm{\nu}}_c^{(k)}) \hat{w}_{c}^{(k)}} {\sum_{c = 1}^{C} p(y_{ij}|\mathbf{x}_{ij},\hat{\bm{\xi}}_{c}^{(k)})\phi(\mathbf{u}_{ij}|\hat{\bm{\mu}}_{c}^{(k)},\hat{\bm{\Sigma}}_c^{(k)}) \psi(\mathbf{v}_{ij}|\hat{\bm{\lambda}}_{c}^{(k)}) \zeta(\bm{d}_{ij}| \hat{\bm{\Gamma}}_c^{(k)},\hat{\bm{\nu}}_c^{(k)}) \hat{w}_{c}^{(k)}}, \label{eq:E-step}
\end{align}
\\
\noindent
which corresponds to the posterior probability that observation $(\mathbf{x}_{ij},y_{ij})$ belongs to the $c-$th cluster, using the current value of
$\bm{\theta}$, i.e., $\hat{\bm{\theta}}^{(k)}$ \citep{Ingrassia2015}. Before advancing to the M-step, we perform the \textit{hard-assignment} step in which we allocate each observation to the cluster with the highest probability. After the E-step, we generate a matrix $\textbf{T} \in \mathbb{R}^{N \scalebox{0.7}{$\times$} C}$ where each row encapsulates the probabilities $\tau_{ijc}$ of observation $i$ within group $j$ belonging to cluster $c$, for $c=1,\ldots, C$. In the hard-assignment step, we define the quantity $\tilde{z}_{ijc} = \operatorname*{argmax}_{c = 1,\ldots,C} \ \tau_{ijc}$ that assigns each observation $ij$ to cluster $c$.

\subsection{M-step}
\label{subsec: M-step}
During the M-step, in the $(k+1)-$th iteration, the goal is to maximize the conditional expectation of $l((\mathbf{x},\mathbf{y}, \mathbf{z}); \bm{\theta})$ given the observed data, in which $z_{ijc}$ is replaced by $\tilde{z}_{ijc}^{(k)}$. See Web Appendix B for the derivation of all formulas in the M-step and a comprehensive pseudo-code of the algorithm.

\subsection{Technical details on the Model}
\label{subsec: selection and prediction}
To obtain a reliable estimate through the algorithm, for each potential number of latent clusters C, we initialize it with completely random assignments of observations to latent clusters. This initialization process is repeated various times, and we retain the iteration with the highest maximized log-likelihood. Finally, the best iteration for all C values are then compared using the Bayesian Information Criterion \cite[BIC,][]{Schwarz1978}. 
The BIC for ML-CWMd is obtained as:
\begin{align}
\label{eq:BIC}
  \text{BIC} &= -2 \cdot \ln(L) + k \cdot \ln(N) \notag \\
    k = \underbrace{C \cdot [ 1 + m ]}_{k_{\text{reg}}} \ + \ \underbrace{C \cdot \left[ \frac{p(p+3)}{2} \right]}_{k_{\text{cont}}} \ &+ \ \underbrace{C \cdot \sum_{r = 1}^{q} (k_r - 1)}_{k_{\text{cat}}} \ + \ \underbrace{C \cdot \left[ \frac{l(l+1)}{2} \right]}_{k_{\text{dich}}} \ + \ \underbrace{C - 1}_{k_{\text{weights}}} 
\end{align}
where $\ln(L)$ is the maximized log-likelihood of the model and $m$ is the number of regression parameters. Additionally, the formula for generating a prediction for a new observation with our model, whose complete derivation is detailed in Web Appendix B, is as follows:
\begin{align}
\label{eq:pred 4}
\mathbb{E}[ y | \mathbf{x}; \bm{\theta}] = p( y = 1 | \mathbf{x}; \bm{\theta}) &= \frac{\sum_{c= 1}^{C} \phi(\mathbf{u}|
\bm{\mu}_{c},\bm{\Sigma}_c) \psi(\mathbf{v}|\bm{\lambda}_{c}) \zeta(\bm{d}| \bm{\Gamma}_c,\bm{\nu}_c)  w_{c} \cdot p(y = 1|\mathbf{x},\bm{\xi}_{c})}{\sum_{c= 1}^{C} \phi(\mathbf{u}|
\bm{\mu}_{c},\bm{\Sigma}_c) \psi(\mathbf{v}|\bm{\lambda}_{c}) \zeta(\bm{d}| \bm{\Gamma}_c,\bm{\nu}_c)  w_{c}}
\end{align}

\section{Simulation Study}
\label{sec: sim study expl}
In this section, we undertake a simulation study to comprehensively assess the performance of the proposed model. The Data Generating Process (DGP) can vary, within each cluster, based on the parameters of covariate distributions, as well as the coefficients of fixed effects and the variance of random effects in the mixed-effects regression component of the model. We consider a DPG composed by $N = 2000$ observations in $J = 10$ groups, each including $n_j = 200$ observations and $C = 3$ latent clusters. The proportion of observations within each cluster is described by parameter $\bm{w}$ = $(0.2, 0.3, 0.5)$. The covariates include: two continuous covariates ($\bm{x}_1$, $\bm{x}_2$) sampled from a multivariate Gaussian of parameters $\bm{\mu}_c$ and $\bm{\Sigma}_c$; two categorical covariates ($\bm{a}_1$, $\bm{a}_2$), with two and three categories respectively, sampled from a multinomial distribution with parameters $\bm{\lambda}_{c1}$, $\bm{\lambda}_{c2}$; and three dichotomous dependent covariates ($\bm{d}_1$, $\bm{d}_2$, $\bm{d}_3$) sampled from an Ising model with parameters $\bm{\Gamma}_c$ and $\bm{\nu}_c$. Let $\textbf{X} =[\bm{x}_{1}, \ \bm{x}_{2}, \ \bm{a}_{1}, \ \bm{a}_{21}, \ \bm{a}_{22}, \ \bm{d}_{1}, \ \bm{d}_{2}, \ \bm{d}_{3}]$ be the design matrix and, therefore, $\bm{x}_{ij}$ be the vector of covariates associated to observation $i$ in group $j$. The linear predictor $\eta_{ijc}$ for observation $i$ in group $j$ and cluster $c$ is defined as follows:
\begin{align}
\label{eq:fixed effects C = 3}
\eta_{ij1} &= \bm{\beta}_{1} \bm{x}^{T}_{ij} + b_{j,1} \ ; \quad \bm{\beta}_{1} = [-0.52, \ 0.08, \ 1.31, \ 0.22, \ 5.33, \ 2.75, \ 2.29, \ 0.93] \notag \\ 
\eta_{ij2} &= \bm{\beta}_{2} \bm{x}^{T}_{ij} + b_{j,2} \ ; \quad \bm{\beta}_{2} = [-0.07, \ 0.79, \ -0.46, \ 0.25, \ -3.89, \ -0.63, \ 0.63, \ -1.51] \\ 
\eta_{ij3} &= \bm{\beta}_{3} \bm{x}^{T}_{ij} + b_{j,3} \ ; \quad \bm{\beta}_{3} = [-0.42, \ -0.31, \ -1.33, \ -0.60, \ -4.18, \ 4.89, \ 3.34, \ -0.46] \notag
\end{align} 
where, $b_{j,c} \sim \mathcal{N}(0,\sigma^2_{b_{c}})$ represents the random intercept associated to group $j$ in cluster $c$ and $\bm{\beta}_{c}$ represents the vector of fixed effects in cluster $c$. 
During each iteration of the simulation process, observations from different groups are randomly assigned to the three clusters ensuring that each cluster contains observations from all groups. Moreover, making use of the same DGP, we generate a test set comprising of 200 observations. The considered true values of the parameters involving the afore-mentioned DGP are detailed in Table~\ref{table 3 cl}.
\begin{table}[H]
\caption{True parameter values pertaining to the Data Generating Process (DPG) of the simulation study across the three considered latent clusters.}
\centering 
    \begin{tabular}{c c c c c c c c c c}
    \hline
    \textbf{Parameter} & \textbf{Cluster 1} & \textbf{Cluster 2} & \textbf{Cluster 3}   \rule[-0.4cm]{0cm}{1cm} \\
    \hline \hline
    $\bm{\mu}_c$ & $(2.05; \ 0.13)$ & $(5.06; \ 4.84)$ & $(4.22; \ -4.51)$   \rule[-0.4cm]{0cm}{1.1cm} \\
    $\bm{\Sigma}_c$ & $\begin{bmatrix}
                0.7 & 0.5 \\
                0.5 & 3.0 
                \end{bmatrix}$ & $\begin{bmatrix}
                2.0 & -1.0 \\
                -1.0 & 3.0 
                \end{bmatrix}$ & $\begin{bmatrix}
                3.0 & 1.0 \\
                1.0 & 2.0
                \end{bmatrix}$  \rule[-1cm]{0cm}{2cm}\\
    $\bm{\lambda}_{c1}$ & $(0.75; \ 0.18; \ 0.07)$ & $(0.07; \ 0.75; \ 0.18)$ & $(0.30; \ 0.51; \ 0.19)$ \rule[-0.4cm]{0cm}{0.6cm} \\
    $\bm{\lambda}_{c2}$ & $(0.51;\ 0.49)$ & $(0.49; \ 0.51)$ & $(0.47; \ 0.53)$ \rule[-0.4cm]{0cm}{1cm}  \\
    $\bm{\gamma}_{l,k}^c$ & $\begin{bmatrix}
                0 & 0.21 & -1.1 \\
                0.21 & 0 & 0 \\
                -1.1 & 0 & 0
                \end{bmatrix}$ & $\begin{bmatrix}
                0 & -0.73 & 0.83 \\
                -0.73 & 0 & 0 \\
                0.83 & 0 & 0
                \end{bmatrix}$ & $\begin{bmatrix}
                0 & -4.15 & 2.11 \\
                -4.15 & 0 & 1.14 \\
                2.11 & 1.14 & 0
                \end{bmatrix}$  \rule[-1cm]{0cm}{2cm}\\
    $\nu_{l}^c$ & $(0.11; \ 0.38; \ -0.49)$ & $(-0.15; \ 0.88; \ -0.18)$ & $(0.73; \ -0.23; \ 0.01)$ \rule[-0.4cm]{0cm}{1cm} \\
    $b_{j,c}$ & $\mathcal{N}(0,4)$ & $\mathcal{N}(0,4)$ & $\mathcal{N}(0,4)$ \rule[-0.4cm]{0cm}{1cm} \\
    \hline
    \end{tabular}
    \label{table 3 cl}
\end{table}

\subsection{Simulation setting and results}
\label{subsec: simulation study C = 3}
\begin{figure}
\centerline{%
\includegraphics[width=130mm]{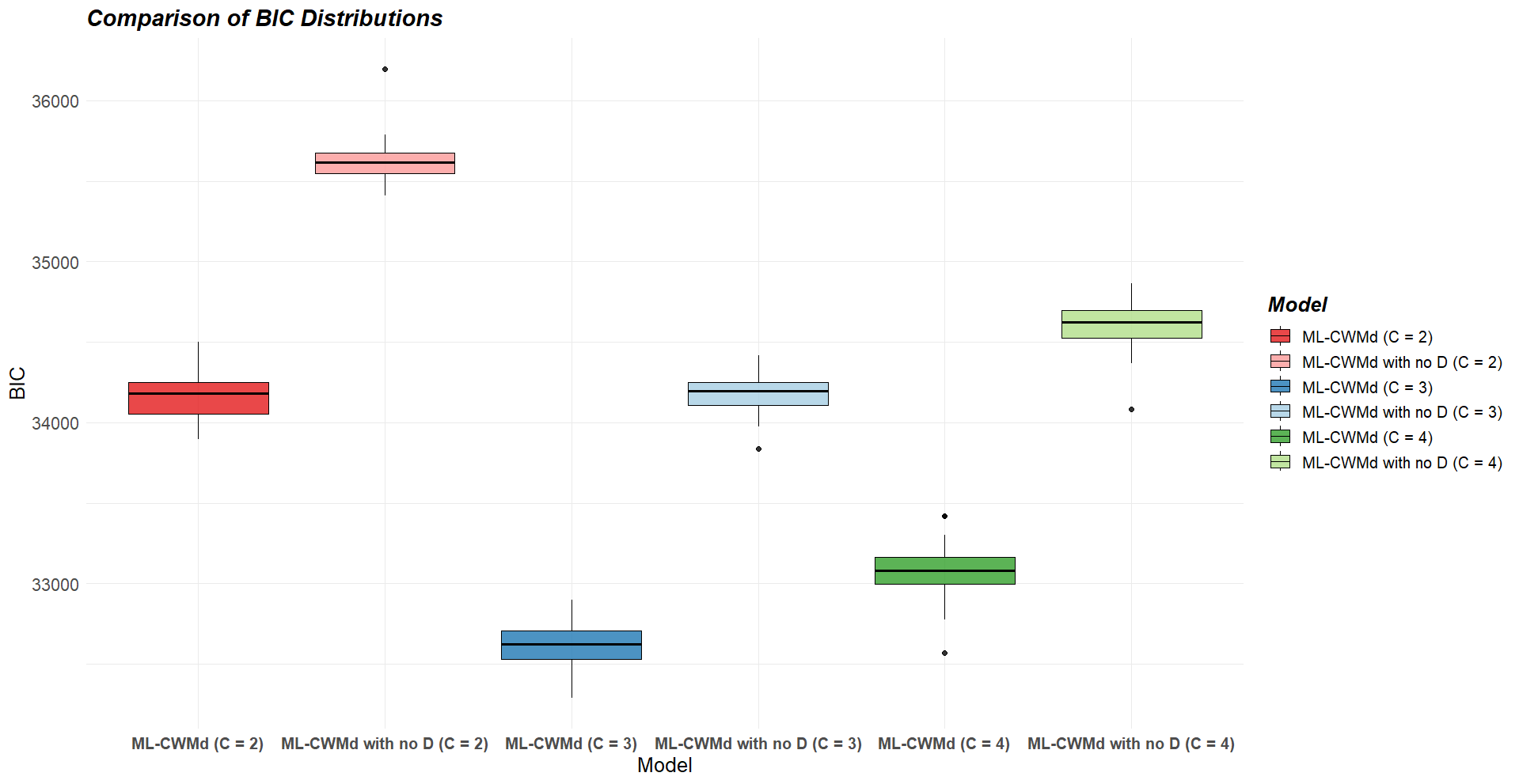}}
\caption{Boxplots of BIC score, defined in Section ~\ref{subsec: selection and prediction}, across the 100 simulations for the ML-CWMd and the ML-CWMd without dichotomous dependent covariates performed with C = 2, 3, 4 clusters.}
\label{BIC-comp}
\end{figure}

\begin{figure}
\centerline{%
\includegraphics[width=80mm]{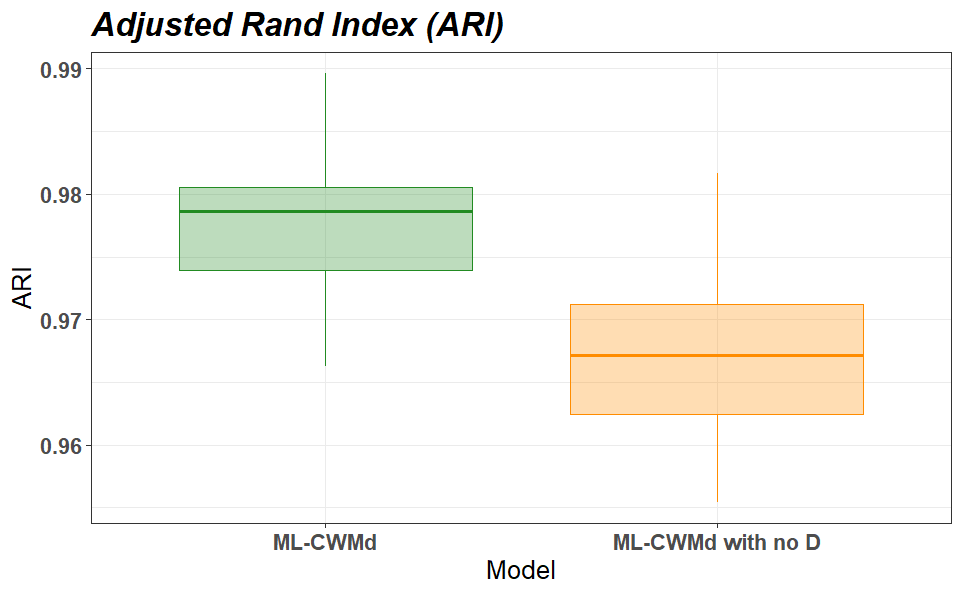}}
\caption{Boxplots of the ARI for the 100 simulations of the simulated experiment. For each method, the number of components C are selected according to the BIC criterion defined in Section ~\ref{subsec: selection and prediction}.}
\label{fig:ari-comp}
\end{figure}

\begin{figure}
    \centering
    \begin{subfigure}[b]{0.45\textwidth}
        \centerline{
        \includegraphics[width=\textwidth]{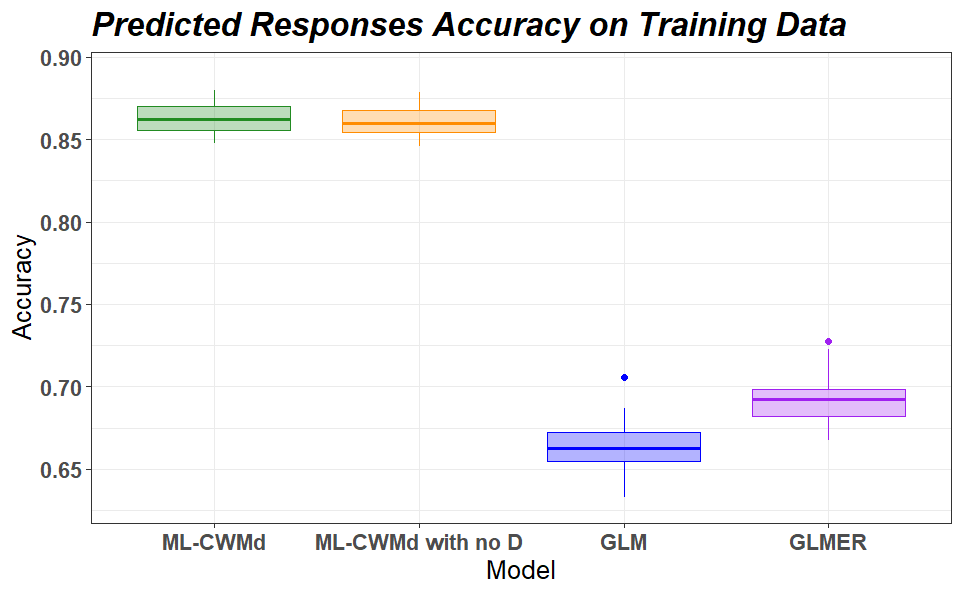}}
        \caption{Boxplots of predicted responses accuracy on the training dataset across the 100 simulation runs for all the models: ML-CWMd, ML-CWMd without dependent binary covariate, GLM and GLMER.}
        \label{fig:acc-comp}
    \end{subfigure}
    \hfill
    \begin{subfigure}[b]{0.45\textwidth}
        \centerline{
        \includegraphics[width=\textwidth]{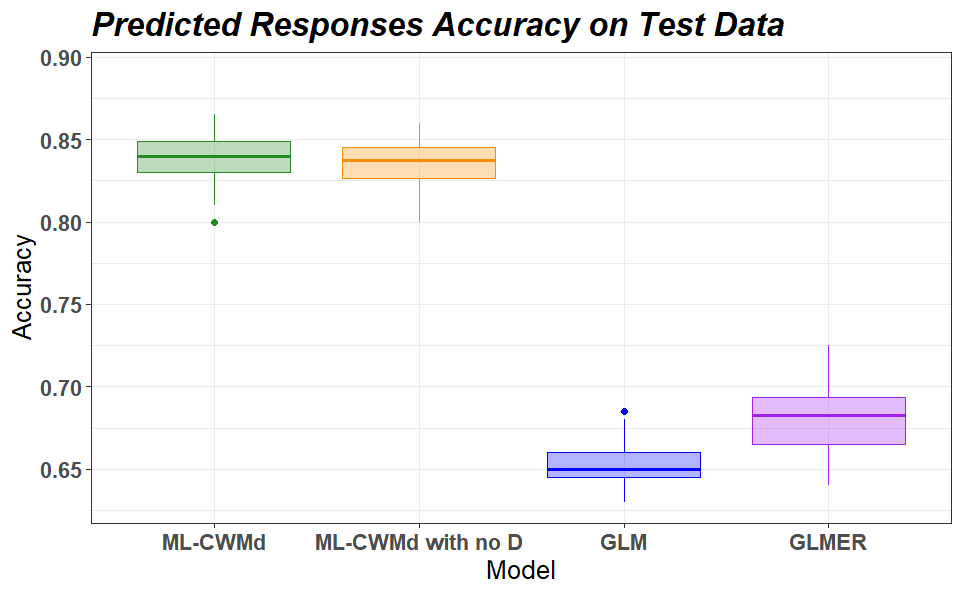}} 
        \caption{Boxplots of predicted responses accuracy on the test dataset across the 100 simulation runs for all the models: ML-CWMd, ML-CWMd without dependent binary covariate, GLM and GLMER.}
        \label{fig:acc-comp-test}
    \end{subfigure}
    \caption{}
    \label{fig:acc-comp-all}
\end{figure}

\begin{figure}
\centerline{%
\includegraphics[width=170mm]{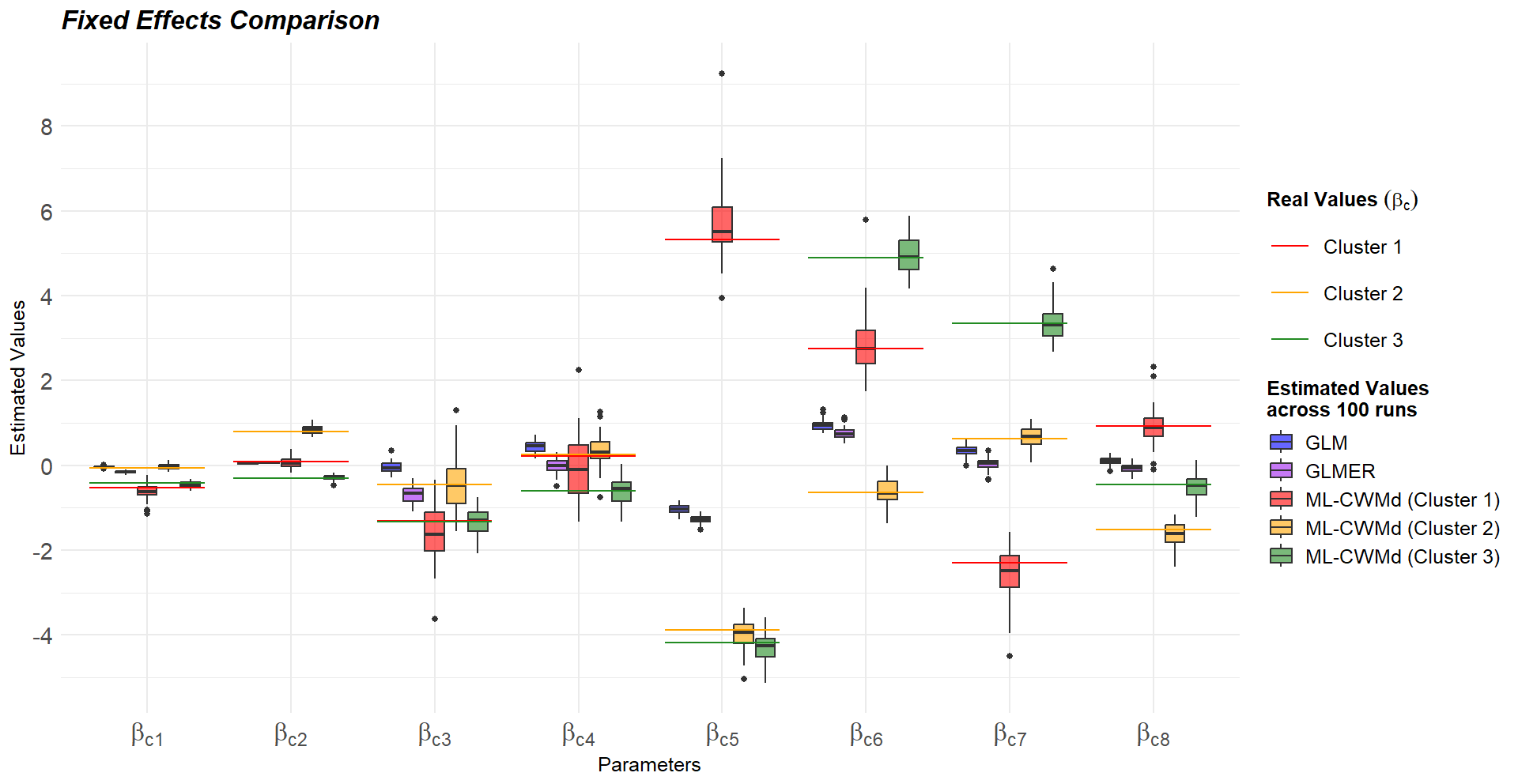}}
\caption{Comparison of the distributions of estimated fixed effects for the 100 runs of the simulated experiment with the true values pertaining to the ML-CWMd (cluster 1 in red, cluster 2 in orange and cluster 3 in green), GLM (blue) and GLMER (purple).}
\label{beta-comp}
\end{figure}

The simulation experiment involves the generation of $100$ synthetic datasets from the DGP described above. Across these simulations, we aim at evaluating the performance of four competing models:
\begin{enumerate}
        \item \textbf{ML-CWMd}: the novel methodology introduced in Section~\ref{s:model} where the dependence among dichotomous covariates is accounted for by means of the Ising model;
        \item \textbf{ML-CWMd with no D}: in this method, differently from ML-CWMd, the binary covariates are modeled using independent binomial distributions akin to categorical covariates with two categories. In this case, the covariates setting follows the one introduced in \cite{Ingrassia2015} and aligns with the methodology of \cite{Berta2019};
        \item \textbf{Generalized Linear Model} (GLM): a logistic regression model for modelling binary responses \citep{dunn2018generalized}.
        \item \textbf{Generalized Linear Mixed Effects Model} (GLMER): a logistic regression model for modelling binary responses  and hierarchical data  \citep{pinheiro2006mixed}.
\end{enumerate}
For what concerns the CWM procedures, for each repetition of the simulated experiment we fit the models by varying $C \in \{2,3,4\}$. Firstly, we aim to assess whether the BIC provides a reliable tool for model selection, identifying the correct DGP and the true number of clusters $C=3$. Figure~\ref{BIC-comp} illustrates the distribution of BIC across 100 simulation runs indicating that the ML-CWMd with Ising and $C = 3$ has the lowest BIC value on average, resulting to be the preferred choice.
Second, we seek to evaluate the accuracy of both models in correctly clustering observations by monitoring the Adjusted Rand Index \citep[ARI,][]{Hubert1985}. As depicted in Figure~\ref{fig:ari-comp}, by efficiently capturing the dependence among the dichotomous covariates, ML-CWMd better recovers the true underlying data partition.
Regarding the other methods, we aim to compare them with the ML-CWMd in terms of predicted responses accuracy on both the training and the test set. For all four models, the optimal classification cutoff is determined through the ROC curve. As depicted in Figure~\ref{fig:acc-comp}, for the training sets, the two ML-CWMd versions exhibit higher accuracy in comparison to GLM and GLMER, with the ML-CWMd with Ising displaying slightly better performance. Employing the identified cutoff from the training data, we calculate the accuracy of predicted responses on the test datasets. As shown in Figure~\ref{fig:acc-comp-test}, the results agree with those displayed on the training set, confirming the superiority of our proposal also in predicting the outcome of previously unseen statistical units. 
Finally, in Figure~\ref{beta-comp}, the distributions of fixed effects $\bm{\beta}_c \ ( \text{for } c = 1, 2, 3)$ recovered from ML-CWMd, GLM, and GLMER across the 100 runs are presented, juxtaposed with their true values. Since ML-CWMd exhibits superior performance compared to ML-CWMd without dependence, and considering the similarity in the distributions of the recovered parameters, we solely display results for the former. ML-CWMd effectively recovers the true parameters; conversely, the estimates derived from GLM and GLMER models often deviate significantly from the true values, especially for coefficients that exhibit considerable variation across clusters, positioning themselves on the average values between clusters.
For further details about the simulation study, Web Appendix C reports the distributions of all estimated parameters for the ML-CWMd across the $100$ repetitions of the simulated experiment, comparing them with their true values.


\section{ML-CWMd for profiling COVID-19 Heart Failure patients and evaluating health providers} 
\label{s: application}
The proposed ML-CWMd model is now employed to profile Heart Failure (HF) patients\footnote{A patient is classified as HF if their records indicate hospitalizations for HF (ICD-9-CM: 428.*) or HF related to hypertension (ICD-9-CM: 402.01, 402.11, and 402.91) in the principal or secondary diagnoses, or under DRG code 127 for HF and shock.} affected by COVID-19 and evaluate the heterogeneity of healthcare providers effects on different risk groups. We focus on HF patients who have been hospitalized due to COVID-19 and who appear in the healthcare administrative database of Lombardy, Italy. 
The purpose of this study is twofold. On one hand, our aim is to enhance comprehension of the characteristics and requirements of HF patients in the region, particularly amidst the exceptional circumstances following the onset of the COVID-19 pandemic. On the other hand, we aim to investigate the capabilities of the proposed model in capturing heterogeneous dynamics within hierarchical data. In details, by applying the ML-CWMd model, we firstly explore the presence of latent HF COVID-19 patient subpopulations, hereafter called \textit{patients profiles}, that display unique sets of attributes and intrinsic characteristics. Secondly, we evaluate and rank hospitals based on their performance in treating patients with different profiles. Third, we deepen our understanding on the association between respiratory illnesses and HF COVID-19 patients death risk, across different patient profiles.
This application provides valuable insights into the complex interplay between patient characteristics, hospital dynamics, and the impact of respiratory illnesses in the context of COVID-19 hospitalizations, ultimately resulting in actionable margins for defining healthcare interventions to enhance the territorial management of patients with HF. 


\subsection{Dataset Description}
\label{dat descr}
\begin{figure}
\centerline{%
\includegraphics[width=170mm]{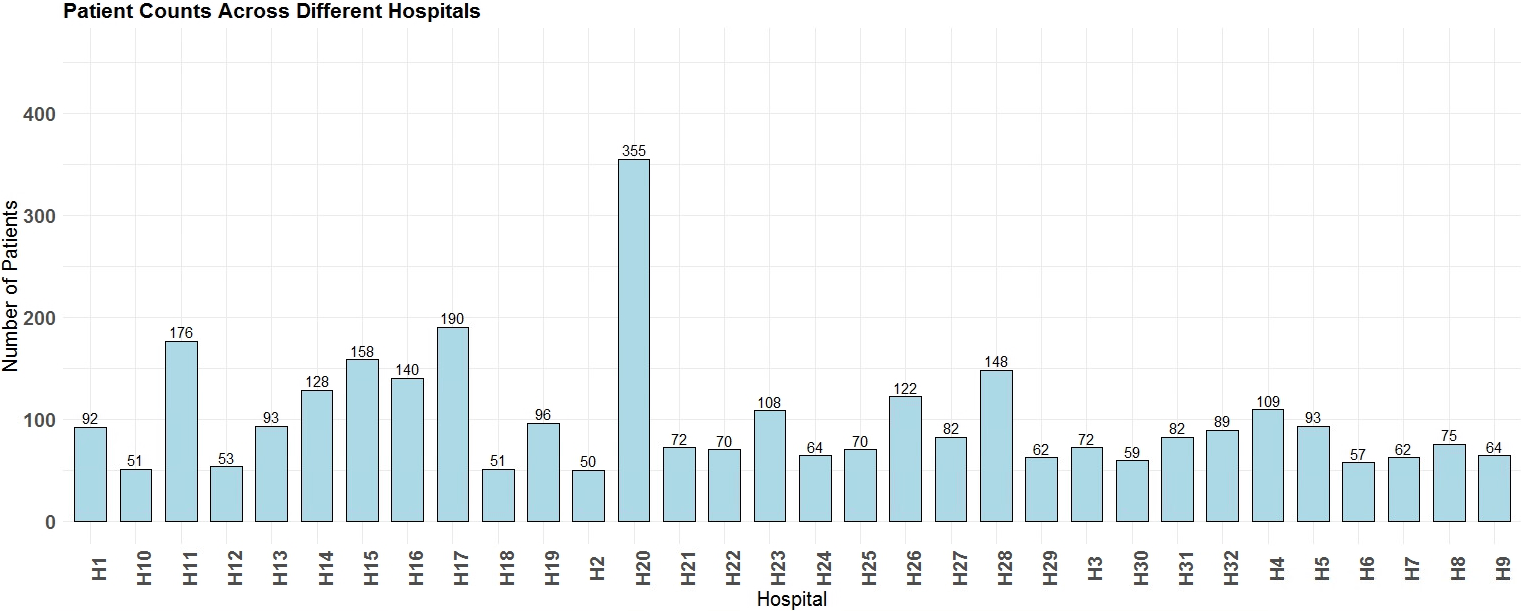}}
\caption{Number of patients within each of the 32 hospitals considered. Only hospital with more than 50 patients treated are considered.}
    \label{fig:App hospitals count}
\end{figure}

From the healthcare administrative records of the  \href{https://www.regione.lombardia.it/wps/portal/istituzionale/HP/DettaglioServizio/servizi-e-informazioni/Enti-e-Operatori/sistema-welfare/Accreditamento/accesso-db-covid/accesso-db-covid}{Lombardy region}, we extract a dataset encompassing $3193$ HF patients who were hospitalized due to COVID-19 across $32$ distinct hospitals in Lombardy region between January 31, 2020 and June 18, 2021. Hospitals with less than $50$ cases were excluded (see Figure~\ref{fig:App hospitals count}). 

\begin{table}[htbp]
\caption{A detailed overview of the descriptive statistics concerning both continuous and categorical variables within the dataset.} 
\renewcommand{\arraystretch}{1.3}
\centering
\begin{tabular}{lccccc c cccc}
 \cline{1-7}
   \textbf{Continuous Variable} && \textbf{Mean}   & \textbf{Min}  & \textbf{Max} & \textbf{Std.Dev} & \textbf{IQR}  \\
\cline{1-7}
\textbf{Age}   && 79.85 & 9 & 106  & 10.28 & \{75; 87\}     \\
\textbf{MCS}   && 10.28 & 0 & 57  & 7.51 & \{5; 14\}    \\
 \cline{1-7}
\multicolumn{1}{c}{} && \multicolumn{5}{c}{} \\
\hhline{----} 
\textbf{Categorical Variable} &&   \multicolumn{1}{@{}l}{\textbf{Categories}} &  \textbf{Percentage}     \\
 \hhline{----} 
 \rowcolor{gray!20}  \textbf{Deceased} && \multicolumn{1}{@{}l}{Yes} &  34.3\%  \\ 
  \rowcolor{gray!5}       &&   \multicolumn{1}{@{}l}{No} &  65.7\%        \\ 
  \hhline{----}
  \rowcolor{gray!20} \textbf{Sex} &&  \multicolumn{1}{@{}l}{Male} &  57.2\% \\ 
    \rowcolor{gray!5}     &&   \multicolumn{1}{@{}l}{Female} &  43.8\%        \\ 
  \hhline{----}
  \rowcolor{gray!20}  \textbf{COPD} &&  \multicolumn{1}{@{}l}{Present} &  30.8\% \\ 
      \rowcolor{gray!5}   &&  \multicolumn{1}{@{}l}{Not Present} &  69.2\%        \\ 
  \hhline{----}
  \rowcolor{gray!20} \textbf{BRH} && \multicolumn{1}{@{}l}{Present} &  15.9\% \\ 
      \rowcolor{gray!5}   &&   \multicolumn{1}{@{}l}{Not Present} &  84.1\%        \\ 
  \hhline{----}
  \rowcolor{gray!20}  \textbf{PNA} &&  \multicolumn{1}{@{}l}{Present} &  21.8\% \\ 
      \rowcolor{gray!5}   &&   \multicolumn{1}{@{}l}{Not Present} &  78.2\%        \\ 
  \hhline{----}
  \rowcolor{gray!20}  \textbf{RF} &&  \multicolumn{1}{@{}l}{Present} &  16.0\% \\ 
     \rowcolor{gray!5}    &&  \multicolumn{1}{@{}l}{Not Present} &  84.0\%        \\ 
  \hhline{----}
 
\end{tabular}
    \label{table Data Reg}
\end{table}

\begin{figure}
\centerline{%
\includegraphics[width=170mm]{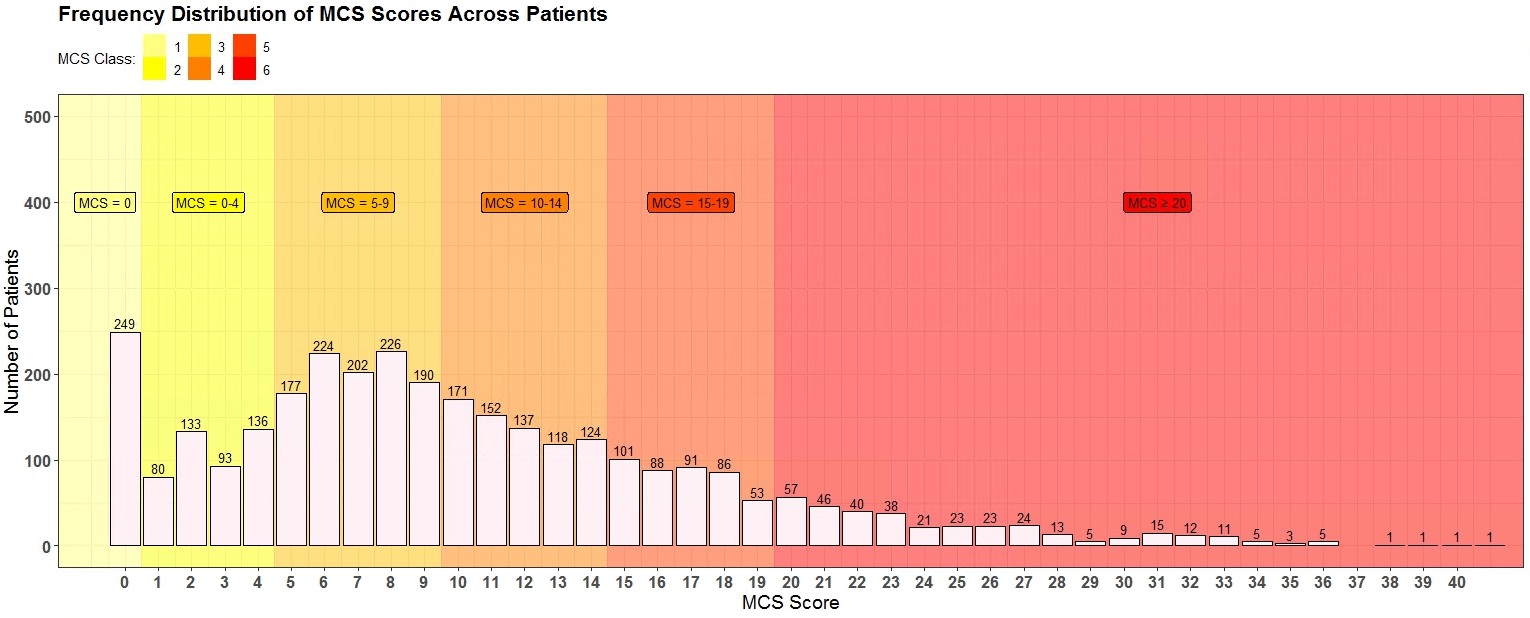}}
\caption{Number of patients within each level of modified MCS categorized across six distinct categories, ranging from 1, signifying no comorbidities, to 6, indicating severe health status.}
    \label{fig:App MCS}
\end{figure}

\noindent
As described in Table~\ref{table Data Reg}, for each patient we observe personal and clinic characteristics:
\begin{itemize}
\item \textbf{Age:} HF patient's age at the moment of COVID-19 hospitalization.
        
\item \textbf{Sex:} gender of the patient.

\item \textbf{Respiratory diseases:} binary variables indicating whether the patient suffers or not from Chronic obstructive pulmonary disease (COPD), pneumonia (PNA), respiratory failure (RF) and bronchitis (BRH) at the moment of admission.
 
        
\item \textbf{Modified Multisource-Comorbidity Score (MCS):} the aim of this scoring system is to offer a succinct snapshot of the patient's clinical condition as described in \cite{corrao2017developing} and \cite{savare2020uso}. We customized the score to the current setting (indicating it as \lq Modified MCS') excluding comorbidities related to respiratory system. In doing so, we aim at evaluating the influence of these particular diseases on a standalone basis with the adjustment in the regression model.  Figure~\ref{fig:App MCS} provides a detailed breakdown of patients' Modified MCS across six distinct categories, reflecting varying degrees of health severity. 
\end{itemize}

\noindent
A more detailed description of the clinical variables and their ICD-9-CM codes \citep{romano1993further} is provided in Web Appendix D. The endpoint we utilize is defined as the mortality of patients within a specific timeframe $t$ following hospitalization. We identify the 45-day threshold as adequate\footnote{We selected this timeframe for two main reasons. Firstly, 89.7\% of the deceased patients passed away within 45 days. Secondly, this timeframe enables us to evaluate both the initial severity of the illness and the potential influence of hospital treatment on patient outcomes. A shorter duration might not offer ample time for the effects of hospital treatment to manifest, while a longer duration could introduce more confounding variables or uncertainties.}, aligning with our research objectives, which include evaluating hospital facilities. Therefore, we created a response (\textit{deceased} in Table~\ref{table Data Reg}) as a dichotomous variable that takes the value of 1 if patient $i$ passed away within $t=45$ days starting from the moment of hospital admission and 0 otherwise. More generally, given the considered time frame $t$ from admission, the regression model will return predictions about the likelihood of a patient's mortality within this specific time frame. This enables the creation of a risk score for each individual patient, representing the probability of death by time $t$.

\subsection{Model Setting and Results}
We utilize the variables age, MCS, gender, COPD and BRH to effectively discern latent patient profiles by including them in the set of random covariates in the CMW framework. Concerning respiratory diseases, we incorporate PNA and RF exclusively into the regression term of the model. The reason for doing so is that COPD and BRH are primary diseases, whereas PNA and RF often result from other underlying conditions. Consequently, we utilize COPD and BRH to assess patient clusters; while examining the impact of PNA and RF on the probability of patient mortality. Regarding the distributions of the covariates, we model age and MCS as continuous variables using a multivariate normal distribution; while, gender, COPD and BRH are modeled as dichotomous dependent covariates using the novel Ising model specification introduced in the proposed ML-CWMd framework. In details, for each patient $i$ in hospital $j$ and for each cluster we consider:
\begin{align}
\text{logit}(\pi_{ij}) &= \alpha_{c} + \beta_{1c} \cdot \text{PNA}_{ij} + \beta_{2c} \cdot \text{RF}_{ij} + b_{cj},
\label{eq: logistic application} 
\end{align}
where $\pi_{ij}$ is the probability of patient to die within 45 days from admission. Following a similar procedure as in Section~\ref{subsec: selection and prediction}, we fit ML-CWMd models varying $C \in$ \{2, 3, 4\}. As reported in Table~\ref{table loglik comp}, the optimal result in terms of BIC is associated with C = 3 clusters.
\begin{table}[htbp]
\renewcommand{\arraystretch}{1.3}
\caption{The outcomes of the procedure outlined in Section~\ref{subsec: selection and prediction} computed for $C \in$ \{2, 3, 4\}. The optimal selection is determined by the lowest BIC, indicating C = 3 as the preferred choice.}
\centering 
\begin{tabular}{c  c  c c}
    \hline
    \textbf{Fixed Clusters} & \textbf{LogLik}  & \textbf{BIC}  \\
    \hline 
     C = 2 & $-2406.1$ & $5570.6$      \\
    \hline
       C = 3 & $-2191.3$ & $5520.2$       \\
    \hline
     C = 4 & $-2117.6$ &  $5752.1$     \\
    \hline
    \end{tabular}
    \label{table loglik comp}
\end{table}

\begin{figure}
    \centering
    \begin{subfigure}[b]{0.3\textwidth}
        \includegraphics[width=\textwidth]{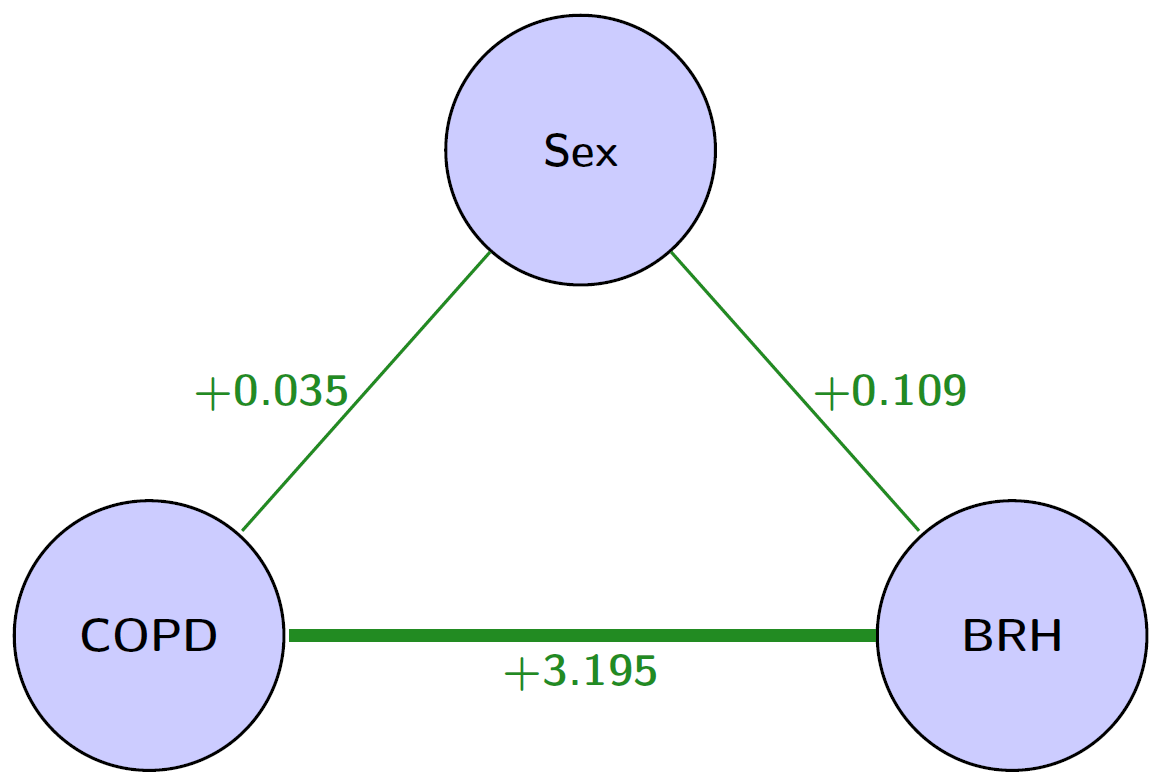}
        \caption{Ising model interaction parameters $\bm{\Gamma}_1$ obtained from ML-CWMd estimation for the first cluster.}
        \label{fig: D1}
    \end{subfigure}
    \hfill
    \begin{subfigure}[b]{0.3\textwidth}
        \includegraphics[width=\textwidth]{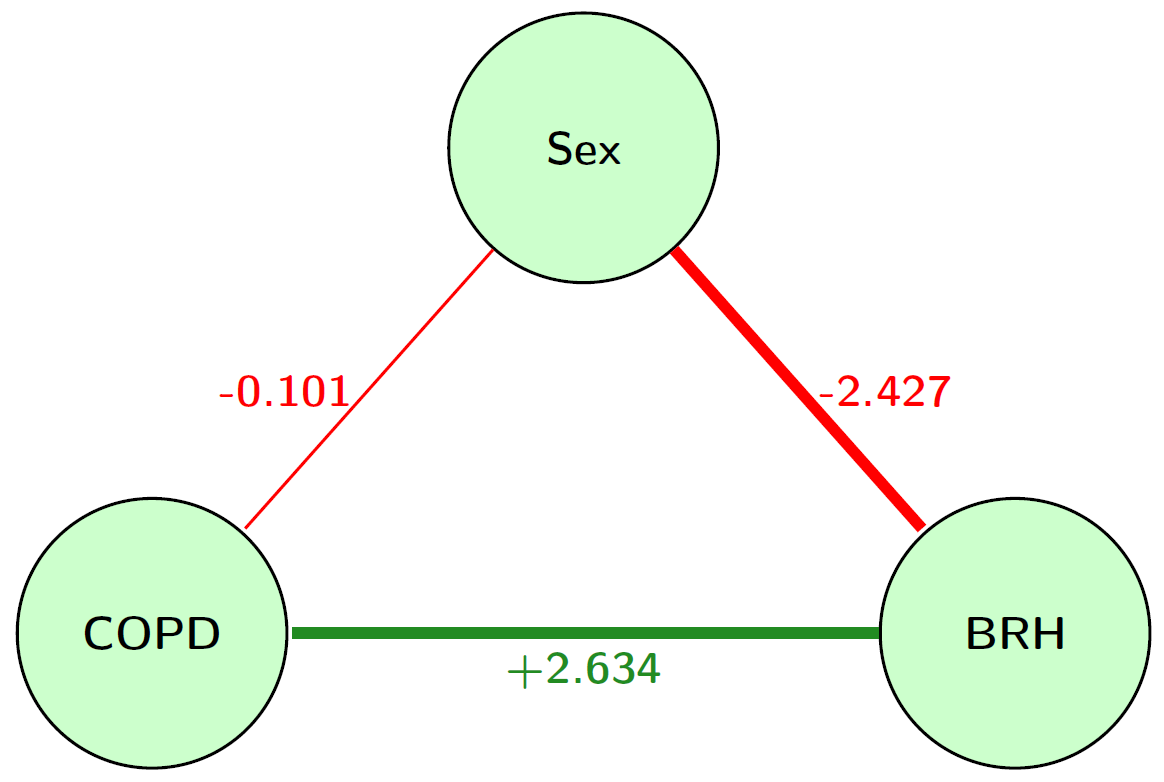} 
        \caption{Ising model interaction parameters $\bm{\Gamma}_2$ obtained from ML-CWMd estimation for the second cluster.}
        \label{fig: D2}
    \end{subfigure}
    \hfill
    \begin{subfigure}[b]{0.3\textwidth}
        \includegraphics[width=\textwidth]{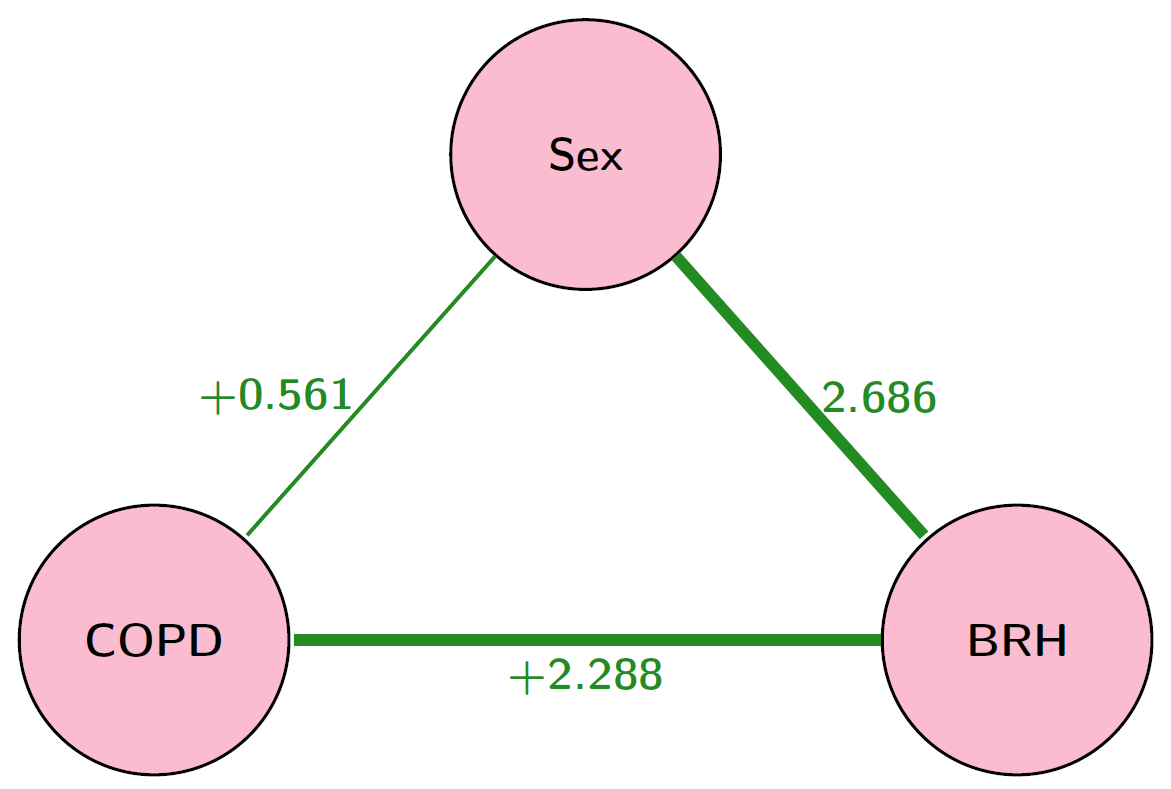}
        \caption{Ising model interaction parameters $\bm{\Gamma}_3$ obtained from ML-CWMd estimation for the third cluster.}
        \label{fig: D3}
    \end{subfigure}
    \caption{}
    \label{fig: D}
\end{figure}

\begin{figure}[htbp]
    \centering
    \begin{subfigure}[b]{1\textwidth}
        \includegraphics[width=\textwidth]{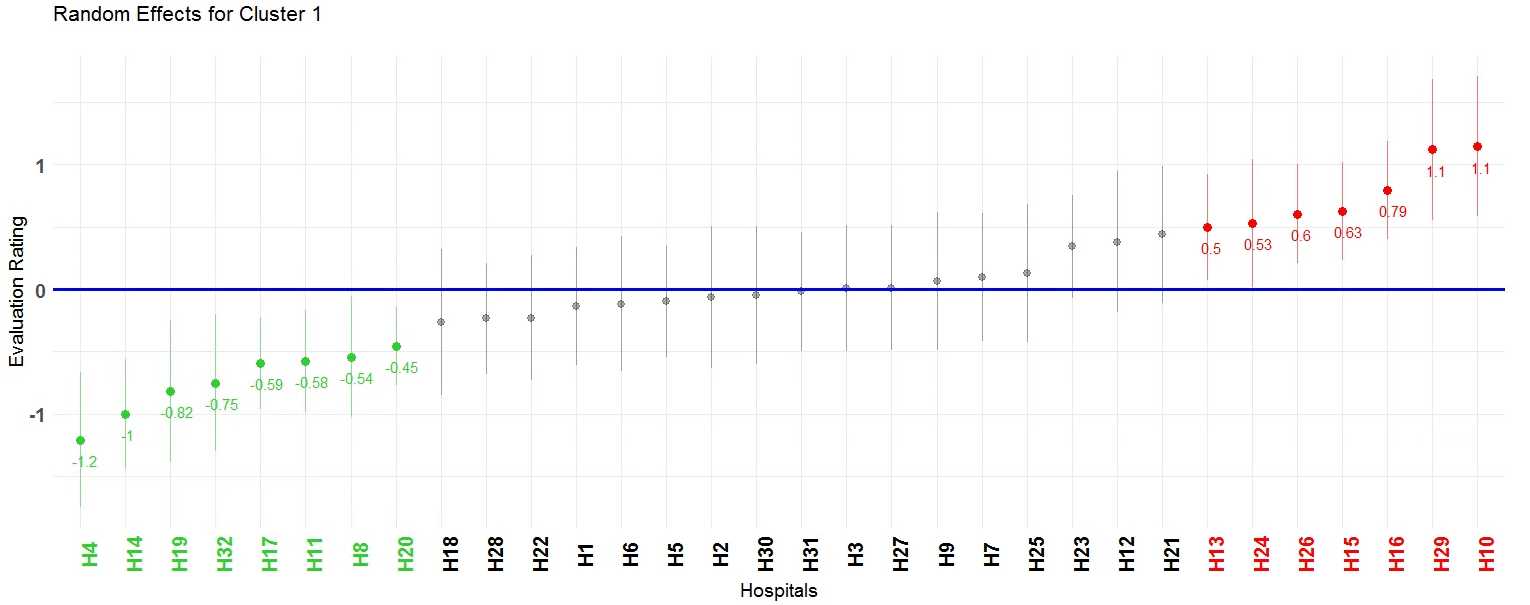}
        \caption{Random effects of the 32 hospitals in the first cluster. Hospitals increasing the probability of death are highlighted in red, while those decreasing it are highlighted in green.}
        \label{fig: H1}
    \end{subfigure}
    \hfill
    \begin{subfigure}[b]{1\textwidth}
        \includegraphics[width=\textwidth]{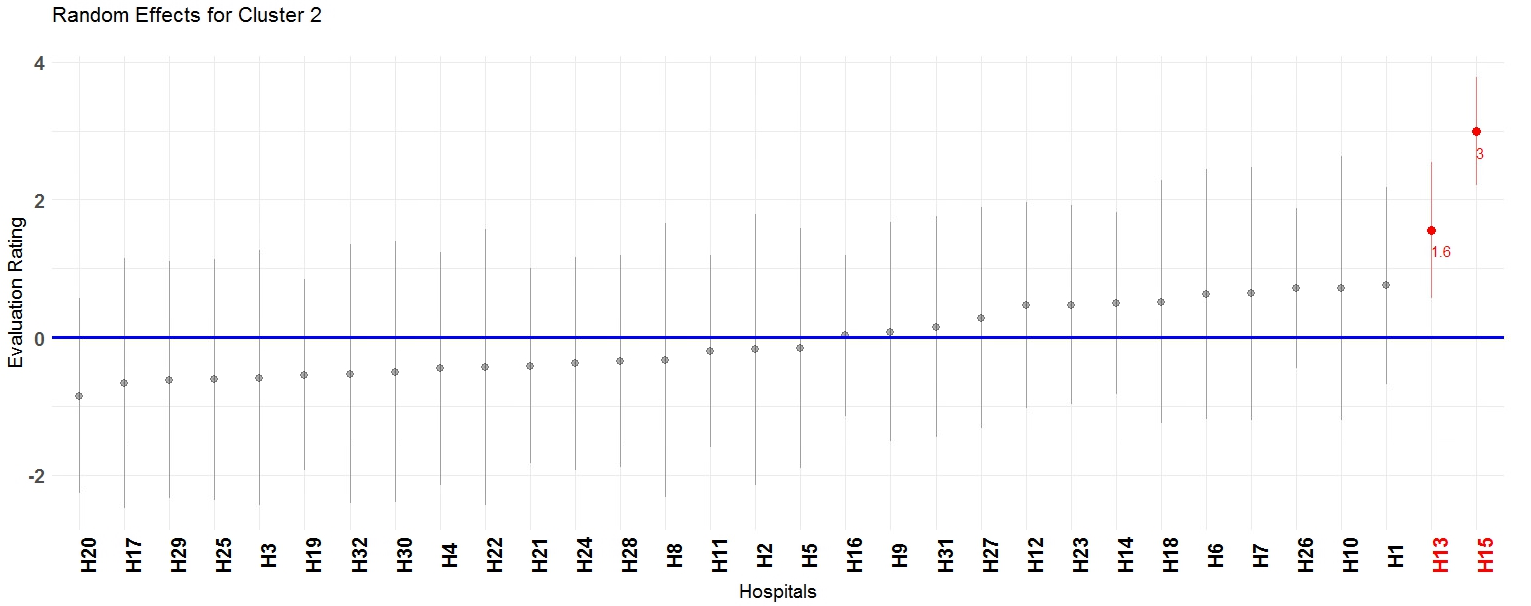} 
        \caption{Random effects of the 32 hospitals in the second cluster. Hospitals increasing the probability of death are highlighted in red, while those decreasing it are highlighted in green.}
        \label{fig: H2}
    \end{subfigure}
    \hfill
    \begin{subfigure}[b]{1\textwidth}
        \includegraphics[width=\textwidth]{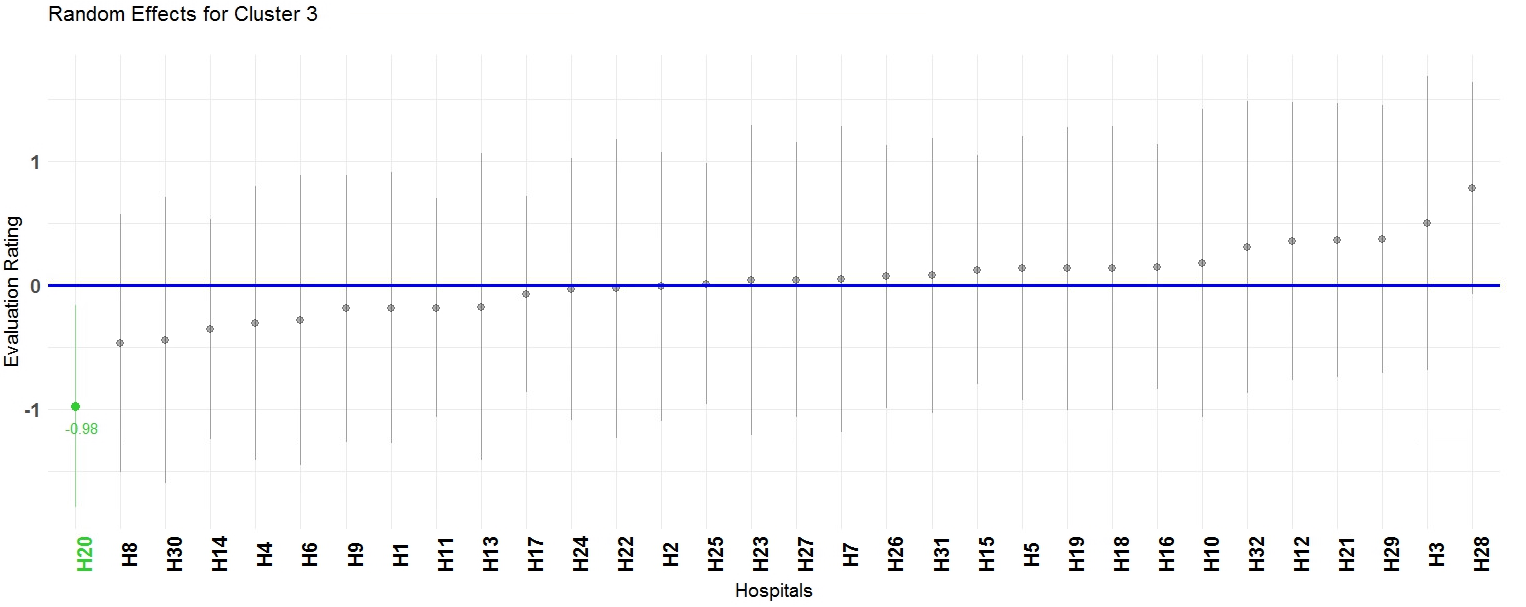}
        \caption{Random effects of the 32 hospitals in the third cluster. Hospitals increasing the probability of death are highlighted in red, while those decreasing it are highlighted in green.}
        \label{fig: H3}
    \end{subfigure}
    \caption{}
    \label{fig: H}
\end{figure}

\noindent
Results are subsequently reported in terms of clusters description, fixed and random effects interpretation and model predictions. The first cluster comprises 2,722 individuals (85.2\% of the total cohort) aged $82.52$ on average with a mean MCS of $9.87$, i.e. individuals with advanced age and a relatively limited occurrence of comorbidities. We label this cluster as the cohort of elderly patients who are in a state of good health. 
Within this cluster, there is gender balance between males (54.6\%) and females (46.4\%), while patients without COPD (67.8\%) and BRH (83.2\%) predominate. As we can see from  Figure~\ref{fig: D1}, the interaction parameters, which signify the relationships between pairs of binary variables, reveal a positive correlation among them all. The interaction between sex and the two respiratory diseases remains minimal, whereas the correlation between COPD and BRH emerges as notably very high. In this cluster, 35.6\% of individuals passed away within 45 days of being admitted. The fixed effects estimates (Table~\ref{tab: fix effect}) reveal that pneumonia is statistically significant, indicating a higher likelihood of mortality among patients with this condition. 
When exploring the random effects related to different hospitals (Figure~\ref{fig: H1}), we discover that 15 out of 32 hospitals demonstrate a significant influence, either in reducing or increasing the likelihood of death among patients in this cluster, with respect to the average. This phenomenon can be attributed to the fact that this patient group comprises elderly individuals who are still in good health, making the appropriateness of hospital instruments and techniques particularly influential. Among the hospitals, those highlighted in green in Figure~\ref{fig: H1} are the ones that significantly lower the probability of death, and thus can be regarded as the most proficient in treating patients belonging to Cluster 1; 
conversely, hospitals highlighted in red are the ones that significantly elevate the probability of death.

\noindent
The second cluster comprises 321 patients aged $59.67$ on average with a mean MCS score of $4.59$. This points towards a cluster composed of young and middle-aged patients who exhibit favorable health conditions. Within this cluster, there is prevalence of males (82.2\%) and a prevalence of patients without COPD (84.4\%) and BRH (93,5\%). Figure~\ref{fig: D2} illustrates the interaction parameters, indicating a notable positive correlation between COPD and BRH, alongside a significant negative correlation between BRH and sex. This suggests a higher prevalence of females with BRH. Within a 45-day period post-hospitalization, the mortality rate among individuals in this cluster stands at 19.3\%. Upon examining the fixed effects (Table~\ref{tab: fix effect}), we observe that the presence of RF in patients within this cluster substantially elevates their risk of mortality. This correlation likely arises from the fact that RF implies impaired lung function, resulting in diminished oxygen levels in the bloodstream. Given that COVID-19 primarily targets the respiratory system, the coexistence of these two conditions further exacerbates respiratory dysfunction, making it challenging for the body to maintain adequate oxygen levels. Consequently, despite their overall health, the severe oxygen deficiency can lead to exceptionally severe complications.
In Figure~\ref{fig: H2}, we observe that only two hospitals 
exhibit significant associations with an increased likelihood of death in these patients.
The limited impact of hospital variability on Cluster 2 is likely due to the fact that patients within this cluster tend to be younger and consistently healthy, making them less vulnerable to the overall quality provided by the treatment facility. 

\noindent
The third cluster comprises a total of 150 patients aged $74.65$ on average along with a mean MCS score of $29.81$. This cluster is predominantly composed of individuals characterized by advanced age and a significant amount of comorbidities. 
Within this cluster, there is prevalence of males (67.5\%) and higher occurence of patients without COPD (62.6\%) and BRH (79.3\%). Figure~\ref{fig: D3} depicts interaction parameters, revealing a substantial positive correlation between COPD and BRH, as well as a significant positive correlation between BRH and gender. This implies a greater prevalence of BRH among males within this cluster. In the context of this cluster, the 43.3\% of individuals have passed away within a 45-day period following hospitalization. RF significantly reduces the probability of death in patients belonging to this cluster (Table~\ref{tab: fix effect}). This somewhat counterituive discovery is quite common among patients of this particular profile \citep{west2014nurse}. Indeed, RF acts as a protective factor, as its presence in patients already facing critical conditions prompts medical doctors to prioritize monitoring and treatment, thereby enhancing their chances of survival. Regarding the random effects (Figure~\ref{fig: H3}), 
only one hospital demonstrates significance in decreasing the probability of death among these patients, with respect to the average. Again suggesting that patients' health conditions are so critical that the specific hospital where they receive treatment has relatively little or no impact on their outcomes.

\begin{figure}
\centerline{%
\includegraphics[width=170mm]{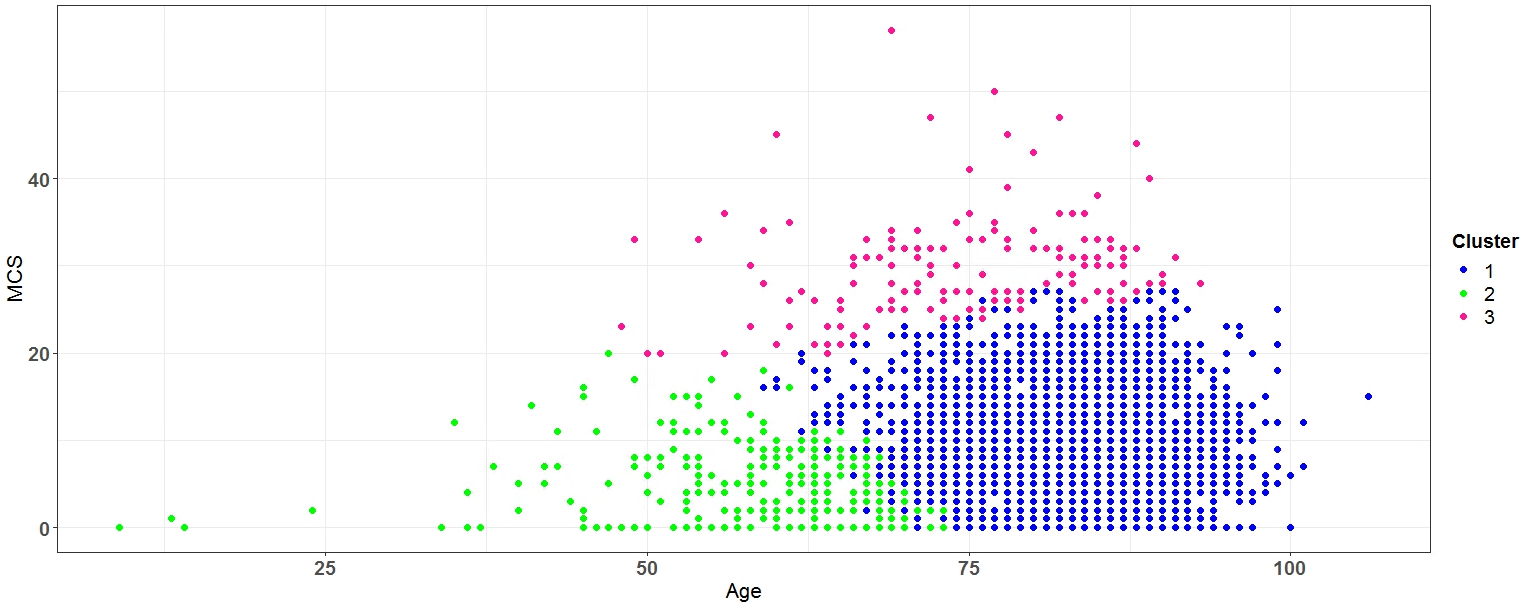}}
\caption{Clusters representation resulting from ML-CWMd estimation on the Age-MCS space.}
\label{fig:Clusters Plot}
\end{figure}

\begin{table}[htbp]
\caption{Values of parameters $\bm{\mu}$, $\bm{\Sigma}$ and $\bm{\nu}$ acquired through ML-CWMd model estimation across the three clusters.}
\centering 
    \begin{tabular}{c  c  c  c  c  c  c}
    \hline
   \textbf{Parameter} &  \textbf{Cluster 1} &  \textbf{Cluster 2} &  \textbf{Cluster 3}   \rule[-0.4cm]{0cm}{1cm} \\
    \hline 
    $\bm{\mu}$ & $(82.52; \ 9.87)$ &  $(59.67; \ 4.59)$ & $(74.65; \ 29.81)$  \rule[-0.4cm]{0cm}{1.1cm} \\
    $\bm{\Sigma}$ & $\begin{bmatrix}
                51.3 & -3.8 \\
                -3.8 & 36.9 
                \end{bmatrix}$ & $\begin{bmatrix}
                91.6 & -6.3 \\
                -6.3 & 19.2 
                \end{bmatrix}$ & $\begin{bmatrix}
                98.6 & 14.7 \\
                14.7 & 35.9
                \end{bmatrix}$   \rule[-1cm]{0cm}{2cm}\\
    $\bm{\nu}$ & $(0.11; \ -3.41; \ -1.37)$ &  $(1.80; \ -2.10; \ -1.90)$ & $(0.28; \ -4.86; \ -1.40)$  \rule[-0.4cm]{0cm}{-2cm} \\
    \hline
    \end{tabular}
    \label{tab: App parameters}
\end{table}

\begin{table}[htbp]
\caption{Estimate of fixed effects parameters $\bm{\beta}_c$ ($c = 1,\hdots,C$) and their corresponding p-values within the three clusters derived from the ML-CWMd estimation process.}
\renewcommand{\arraystretch}{1.3}
\centering
\begin{tabular}{lccccc c ccc ccc}
 \cline{3-4} \cline{6-7} \cline{9-10}
 \multicolumn{1}{c}{ }  && \multicolumn{2}{c}{\textbf{Cluster 1}} && \multicolumn{2}{c}{\textbf{Cluster 2}} && \multicolumn{2}{c}{\textbf{Cluster 3}}  \\
 \cline{1-1} \cline{3-4} \cline{6-7} \cline{9-10}
  \textbf{Variable}  && \textbf{Estimate} & {$\textbf{Pr}(>|z|)$}  && \textbf{Estimate} & {$\textbf{Pr}(>|z|)$}    && \textbf{Estimate} & {$\textbf{Pr}(>|z|)$}  \\
  \cline{1-1} \cline{3-4} \cline{6-7} \cline{9-10}
 \textbf{PNA}  &&  $+0.18$ & $0.09 \ .$  && $-0.08$ & $0.21$    && $+0.27$ & $0.47$   \\
 \textbf{RF}  &&  $+0.10$ & $0.37$  && $+1.51$ & $0.018$*    && $-0.90$ & $0.046$*   \\
\end{tabular}
    \label{tab: fix effect}
\end{table}

When examining the random effects within the various clusters, a noteworthy trend emerges in the assessment of hospitals. For instance, Hospital H20 consistently retrieves positive evaluations in both Clusters 1 and 3, whereas Hospitals 13 and 15 receive unfavorable ratings in both Clusters 1 and 2. This discovery has important implications for policy making allowing suitable monitoring and second level analysis of \lq out of control' situations.
Finally, we fit both a GLMER and a GLM on the same data and, determining the optimal threshold via the ROC curve, we extracted the accuracy of the three models. Table~\ref{tab: pred acc} reveals that ML-CWMd achieves the highest prediction accuracy at 66.5\%.

\begin{table}[H]
\renewcommand{\arraystretch}{1.3}
\caption{Comparison of predicted responses accuracy on the training dataset among three competing models: ML-CWMd, GLMER and GLM.}
\centering 
    \begin{tabular}{l l l}
     \hline
    \textbf{Model} & \textbf{Accuracy}    \\
    \hline 
      ML-CWMd & $0.665$     \\
    \hline
        GLMER & $0.659$      \\
    \hline
     GLM & $0.584$     \\
    \hline
    \end{tabular}
    \label{tab: pred acc}
\end{table}

\subsection{Scenario Analysis}
In this section, we implement a scenario analysis to emphasize the importance of considering both the cluster factor and the hospital effect to capture the real-world complexity.
We define three new COVID-19 HF patients, with characteristics closely mirroring those identified within the three clusters. From each of these patients, we select one with respiratory diseases (PNA and RF), and one without them, resulting in six possible profiles (Table~\ref{table ree stuud}). The objective is to illustrate the difference between predicting the likelihood of mortality for these new patients using the proposed model (Equation~\eqref{eq:pred 4}), a GLMER that neglects the consideration of latent patient clusters, and a GLM that neglects both the consideration of latent patient cluster and hospital effect. For each patient, we delineate three distinct predictions, corresponding, respectively, to random effects equivalent to $-1\hat{\sigma}_{bc}$ (hospital with commendable performance concerning that specific patient profile), $0$ (hospital with negligible influence on the outcome for that specific patient profile) and $1\hat{\sigma}_{bc}$ (hospital with poor performance regarding that specific patient profile).


\begin{table}[htbp]
\renewcommand{\arraystretch}{1.3}
\caption{Visual depiction of the characteristics of the six new patients for whom we calculate predicted probabilities of mortality in the scenario analysis.}
\centering 
    \begin{tabular}{c c c c c c c c c c}
    \hline
    \textbf{Patient ID} & \textbf{Cluster} & \textbf{Age} & \textbf{MCS} & \textbf{Sex} & \textbf{BRH} &  \textbf{COPD} & \textbf{RF} &\textbf{PNA}    \\
    \hline 
    $\text{A}$  &  1 & 83 & 7.6 & M & 1 & 0 & 0 & 0   \\
    \hline
    $\text{B}$  &  2 & 59 & 5.2 & M & 1 & 0 & 0 & 0   \\
    \hline
    $\text{C}$  &  3 & 77.5 & 26.9 & M & 1 & 0 & 0 & 0   \\
    \hline
     $\text{a}$  &  1 & 83 & 7.6 & M & 1 & 0 & 1 & 1   \\
    \hline
    $\text{b}$  &  2 & 59 & 5.2 & M & 1 & 0 & 1 & 1   \\
    \hline
    $\text{c}$ &  3 & 77.5 & 26.9 & M & 1 & 0 & 1 & 1   \\
    \hline
    \end{tabular}
    \label{table ree stuud}
\end{table}

\begin{figure}
\centerline{%
\includegraphics[width=170mm]{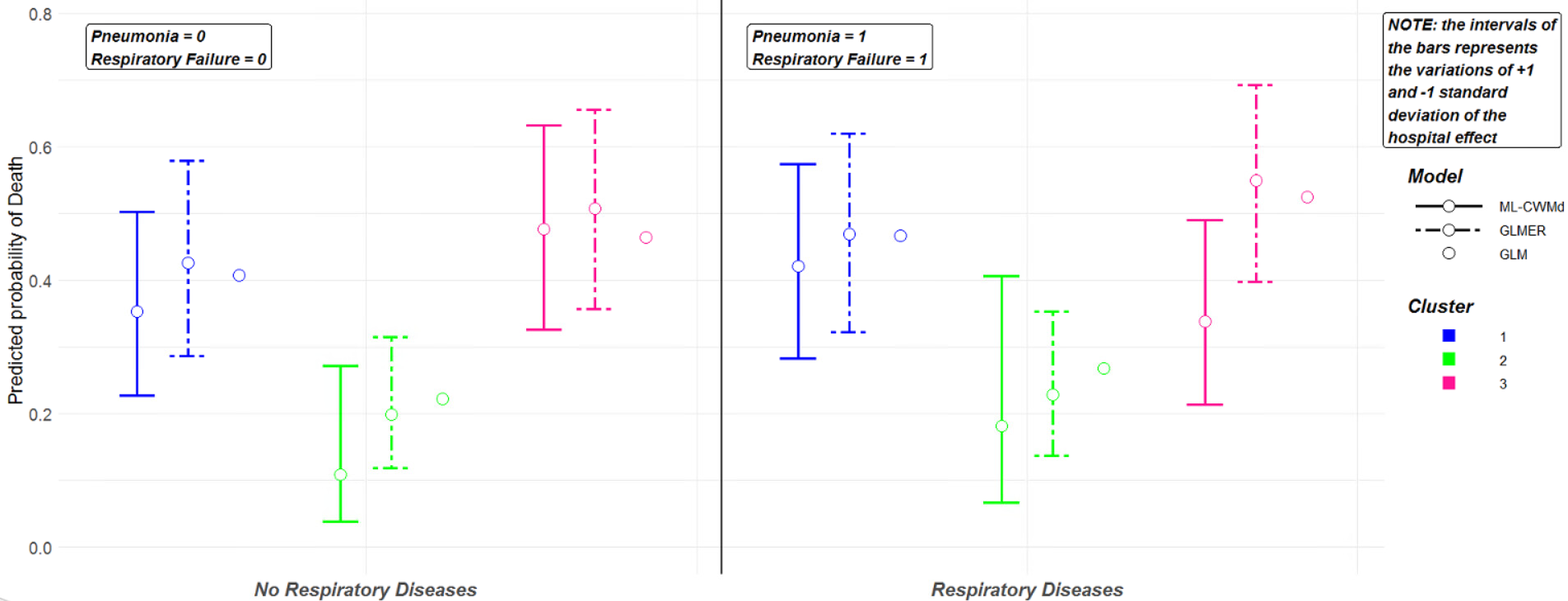}}
\caption{Predicted probabilities of mortality for the six patients across the three competing models. Bar intervals illustrate the shift in predicted probabilities when the hospital is deemed advantageous versus detrimental. Solid bars represent ML-CWMd, dotted bars denote GLMER, and single points indicate GLM.}
\label{fig: scenario}
\end{figure}
In Figure~\ref{fig: scenario}\footnote{For the GLM, we have only a single value since it does not take into account the hospital effect}, we show the predictions obtained with the three models. The bar intervals on the graph show how the predicted probabilities of death change when the hospital is considered beneficial for patient characteristics (at $-1\hat{\sigma}_{bc}$ ) versus when it is considered detrimental (at $+1\hat{\sigma}_{bc}$ ). The GLMER generates highly similar predictions across the same patient with and without respiratory diseases because it does not identify any respiratory diseases as significant covariates. Notably, only age, sex, and MCS emerge as significant variables in the fixed regression component of the GLMER. Consequently, GLMER fails to capture the variability attributed to respiratory diseases, which could significantly influence a patient probability of death, particularly among COVID-19 HF patients. Instead, using the ML-CWMd enable us to take into consideration the respiratory diseases effects observing a significant change in the probability of death based on the presence or absence of these diseases in the ML-CWMd predictions. For instance, for the patient in the green cluster (young individuals with low MCS), the probability of death in a hospital characterized as detrimental for the patient characteristics is approximately 0.28 if the patient has no respiratory diseases. This probability increases to around 0.4 if the patient has both respiratory diseases. In conclusion, the ML-CWMd provides more precise estimations of patient survival probabilities across different risk categories. Additionally, it offers a valid tool for the assessment and monitoring of hospital facilities utilizing administrative databases.

\section{Discussion}
\label{s:discuss}
\color{black}
This paper stands as a novel proposal for jointly risk stratifying patients and assessing their related groupers, serving healthcare management and public health purposes. A novel Multilevel Logistic Cluster-Weighted Model (ML-CWMd) is introduced, extending previous works with the ability to effectively capture the dependence among observations within the same cluster and hierarchy; as well as among dichotomous variables through the inclusion of the Ising model contribution. Although originally designed for medical applications, the ML-CWMd can serve as a valuable tool within other domains, where groups and units encapsulate distinct entities or phenomena. Resorting to maximum likelihood estimation, we have devised a tailored Classification Expectation Maximization algorithm to perform model fitting, testing its performance in a simulated setting, and comparing it with state-of-the-art alternatives. Our proposal has demonstrated promising results when dealing with complex scenarios encompassing latent clusters of observations, group effects, and the interdependence among binary covariates. This research has been spurred by the challenge of developing tailored models adept at accommodating diverse patient profiles and hospital-specific effects. Specifically, we applied our proposal to a real-world administrative dataset of Lombardy region, Italy, including information about Heart Failure patients hospitalized for COVID-19. The analysis has revealed the existence of three distinct patient profiles, each characterized by cluster-wise different survival patterns and comorbidities. On top of this, the model setting allows for valuable insights into the ways respiratory diseases and hospitals impact individual profiles of patients. The analysis has demonstrated promising results in terms of actionable margins for defining necessary healthcare interventions to enhance the territorial management of patients with heart failure through the planning of optimal care pathways, thereby reducing adverse clinical outcomes and improving system efficiency. Of course, the devised methodology also possesses some limitations. The independence assumption among observations belonging to the same known hierarchy but different latent cluster may not always be tenable, as the grouping effect could potentially exhibit shared patterns across clusters. Furthermore, the restriction of modeling dependence among dichotomous covariates only was solely motivated by the type of covariates available within the Lombardy region database. However, there may be a need to enrich our modeling framework by accommodating dependent categorical covariates with more than two categories. A direction for future research involves modifying the current proposal to overcome the aforementioned limitations, as well as extending its capability to model time-to-event outcomes. Several proposals are currently under study, and they will be the focus of future works.

\section*{Acknowledgements}

This work is part of the ENHANCE-HEART project: Efficacy evaluatioN of the therapeutic-care patHways, of
the heAlthcare providers effects aNd of the risk stratifiCation in patiEnts suffering from HEART failure. The authors thank the ‘Unità Organizzativa Osservatorio Epidemiologico Regionale’ and ARIA S.p.A for providing data and technological support. Andrea Cappozzo acknowledges financial support from the Italian Ministry of University and Research (MUR) under the Department of Excellence 2023-2027 grant agreement ``Centre of Excellence in Economics and Data Science'' (CEEDS). The present research is part of the activities of ``Dipartimento di Eccellenza 2023-2027''. The authors gratefully acknowledge the support from the Department of Mathematics of Politecnico di Milano, which facilitated this research as part of the department's ongoing activities. \vspace*{-8pt}

\bibliographystyle{unsrtnat}
\bibliography{library}  

\newpage
\section*{Supplementary Materials}

\section*{Web Appendix A: Further details on the Ising Model}
In this appendix, we provide further details on the
Ising model \citep{Haslbeck2021}. The general interpretation entails that the parameters $\bm{\Gamma}_c$, also referred to as interaction
parameters, depict the interrelationships between the binary variables. Meanwhile, the parameters $\bm{\nu}_c$, also
known as threshold parameters, highlight the inclination of a variable to lean towards one state or the other
when all interaction parameters related to that particular variable are devoid of influence, i.e. equal to zero.
When dealing with a dichotomous variable, we have the option to express the labels of the two classes using
either the \{0,1\} domain or the \{-1,1\} domain. The Ising model has formulations in both of these domains. It
is feasible to convert the parameters of an Ising model from one range to the other, resulting in a statistically
equivalent model. The distinction between the two domains lies in how the interaction parameters are interpreted.
To elucidate the alterations in how interaction parameters are understood across the two domains, we center
our attention on a scenario involving two variables. Within the \{0, 1\} domain, augmenting the interaction
parameter between the two variables, denoted as $\bm{\gamma}_{12}$, results in an increased likelihood of the state (1, 1) when
compared to the other feasible states: (0, 0), (0, 1), and (1, 0). Conversely, within the \{-1, 1\} domain, elevating
the interaction parameter between the two variables, $\bm{\gamma}_{12}$, gives rise to a higher probability of the states (1, 1)
and (-1, -1), in contrast to the states (-1, 1) and (1, -1). Based on the assumed pattern of interdependence
between the dichotomous variables, we can choose to utilize either of the two domains.
The \{0,1\} domain could be more suitable when considering situations where correlation is presumed to exist only
between the presence of both binary variables within the pairs being examined. As an example, when aiming
to model the dependence between the presence or absence of different diseases in patients, using this specific
domain would likely yield more suitable results since it is reasonable to presume that there is no correlation
between the absence of either disease within a pair of the diseases under consideration. However, the possibility
of a connection between the presence of both diseases remains plausible.
Conversely, the \{-1,1\} domain is valuable in scenarios where it is hypothesized that correlation exists between
both the presence of both variables and the absence of both variables within the pairs under scrutiny. To provide a clearer explanation of the Ising model, we illustrate an example concerning the third cluster of the simulation study DPG presented in the main paper, where the parameters of the Ising model are the following: 
\begin{align}
  \bm{\nu}_c = (0.73; \ -0.23; \ 0.01) \quad \quad ; \quad \quad \bm{\Gamma}_c = \begin{bmatrix}
           0 & -4.15 & 2.11 \\
                -4.15 & 0 & 1.14 \\
                2.11 & 1.14 & 0
                \end{bmatrix}
\end{align}
In the case of three dichotomous dependent variables $(\textbf{d}_{1}, \textbf{d}_{2}, \textbf{d}_{3})$, the Ising model assumes the following form: 
\begin{align}
\label{eq:zeta density 2}
\zeta({d}_{1}, {d}_{2}, {d}_{3}) =  \frac{1}{S} \exp \bigg{(} \nu_1 d_1 + \nu_2 d_2 + \nu_3 d_3 +\bm{\gamma}_{12} d_1 d_2 + \bm{\gamma}_{13} d_1 d_3 + \bm{\gamma}_{23} d_2 d_3 \bigg{)}
\end{align}
We model these variables using the \{0,1\} domain. Consequently, after acquiring estimates for the parameters of the ML-CWMd model pertaining to the Ising model (in this instance, we directly utilize their actual values), we can derive, within each cluster, the probabilities associated with every possible combination of states as follows:
\begin{align}
\label{eq:zeta density}
\mathbb{P}({d}_{1} = 0, {d}_{2} = 0, {d}_{3} = 0) &=  \frac{1}{S} \notag \\
\mathbb{P}({d}_{1} = 1, {d}_{2} = 0, {d}_{3} = 0) &=  \frac{1}{S} \exp \big{(} \nu_1  \big{)} \notag \\
\mathbb{P}({d}_{1} = 0, {d}_{2} = 1, {d}_{3} = 0) &=  \frac{1}{S} \exp \big{(} \nu_2  \big{)} \notag \\
\mathbb{P}({d}_{1} = 0, {d}_{2} = 0, {d}_{3} = 1) &=  \frac{1}{S} \exp \big{(} \nu_3  \big{)} \notag \\
\mathbb{P}({d}_{1} = 1, {d}_{2} = 1, {d}_{3} = 0) &= \frac{1}{S} \exp \big{(} \nu_1 + \nu_2  +\bm{\gamma}_{12} \big{)} \notag \\
\mathbb{P}({d}_{1} = 1, {d}_{2} = 0, {d}_{3} = 1) &= \frac{1}{S} \exp \big{(} \nu_1 + \nu_3  +\bm{\gamma}_{13} \big{)} \notag \\
\mathbb{P}({d}_{1} = 0, {d}_{2} = 1, {d}_{3} = 1) &= \frac{1}{S} \exp \big{(} \nu_2 + \nu_3  +\bm{\gamma}_{23} \big{)} \notag \\
\mathbb{P}({d}_{1} = 0, {d}_{2} = 1, {d}_{3} = 1) &= \frac{1}{S} \exp \big{(} \nu_1 + \nu_2 + \nu_3 + \bm{\gamma}_{12} + \bm{\gamma}_{13} + \bm{\gamma}_{23} \big{)} \notag 
\end{align}

\begin{table*}[htbp]
\renewcommand{\arraystretch}{1.3}
\centering
\begin{tabular}{|c | c | c| c|}
    \hline
    \rowcolor{gray!20}
    \textbf{Probability} & $\textbf{d}_{1}$ & $\textbf{d}_{2}$ & $\textbf{d}_{3}$    \\
    \hline \hline
    0.039  &  0 & 0 & 0   \\
    \hline
    0.082  &  1 & 0 & 0    \\
    \hline
    0.031  &   0 & 1 & 0     \\
    \hline
     0.039 &   0 & 0 & 1     \\
    \hline
    0.001  &   1 & 1 & 0    \\
    \hline
    0.682  &   1 & 0 & 1   \\
    \hline
    0.100  &   0 & 1 & 1    \\
    \hline
    0.026 &   1 & 1 & 1    \\
    \hline
    \end{tabular}
\caption{Probabilities related to all combinations of the states of variables $(\textbf{d}_{1}, \textbf{d}_{2}, \textbf{d}_{3})$.}
\label{table ree stuud 2}
\end{table*}
\noindent
The probabilities related to this specific case corresponding to all combinations of states are showed in Table~\ref{table ree stuud}. To determine the strength of correlation between two variables, it is essential to assess their interaction parameter value alongside all other parameters, gauging its magnitude relative to them. This evaluation allows us to discern its influence on the probabilities of all possible states. If its magnitude is comparable to or even greater to the one of other parameters (both interaction and threshold parameters values), we can infer a strong interaction. In the particular scenario being examined, it is evident that the interaction parameters between ($\textbf{d}_1$, $\textbf{d}_2$) and ($\textbf{d}_1$, $\textbf{d}_3$) exhibit significant strength when compared to all other values. Indeed, the probability of the state (1,0,1), which is notably influenced by the magnitude of these parameters, is the highest.

\section*{Web Appendix B: Additional details on the M-step and prediction on a new observation}
\label{comp details M}

In this section, we report details about the maximization step of the tailored EM algorithm.  
At the $k-$th iteration, the functional to be maximized is the following: 
\begin{subequations}
\label{eq:M-step sup}
\begin{align}
Q(\bm{\theta},\bm{\theta}^{(k)}) 
= \sum_{c= 1}^{C}  \sum_{j = 1}^{J}  \tilde{\bm{z}}_{jc}^{(k)}((\mathbf{x},y), \hat{\bm{\theta}}^{(k)}) \log\big{(}p(\mathbf{y}_{j}|\mathbf{x}_{j},\bm{\xi}_{c})\big{)}   \ +
\label{eq: M-step regression sup} \\
+ \ \sum_{c= 1}^{C}  \sum_{j = 1}^{J} \sum_{i = 1}^{n_j}  \tilde{z}_{ijc}^{(k)}((\mathbf{x},y), \hat{\bm{\theta}}^{(k)})  \log\big{(}\phi(\mathbf{u}_{ij}|\bm{\mu}_{c},\bm{\Sigma}_c)\big{)} \ +
\label{eq: M-step continuous sup} \\
 + \ \sum_{c= 1}^{C}  \sum_{j = 1}^{J} \sum_{i = 1}^{n_j}  \tilde{z}_{ijc}^{(k)}((\mathbf{x},y), \hat{\bm{\theta}}^{(k)}) \log\big{(}\psi(\mathbf{v}_{ij}|\bm{\lambda}_{c})\big{)} \ + 
  \label{eq: M-step categorical} \\
+ \  \sum_{c= 1}^{C}  \sum_{j = 1}^{J} \sum_{i = 1}^{n_j} \tilde{z}_{ijc}^{(k)}((\mathbf{x},y), \hat{\bm{\theta}}^{(k)}) \log\big{(}\zeta(\bm{d}_{ij}| \bm{\Gamma}_c,\bm{\nu}_c)\big{)} \ + 
\label{eq: M-step dich dep sup} \\
 + \ \sum_{c= 1}^{C}  \sum_{j = 1}^{J} \sum_{i = 1}^{n_j}  \tilde{z}_{ijc}^{(k)}((\mathbf{x},y), \hat{\bm{\theta}}^{(k)})  \log\big{(}w_{c}\big{)}.
 \label{eq: M-step w}
\end{align}
\end{subequations}
As there are no cross-derivatives between the five terms, their maximization can be carried out separately.

\subsection{Maximization of the Regression component}
\label{subsubsec: max first term}
The maximization of $Q(\bm{\theta},\bm{\theta}^{(k)})$ with respect to $\bm{\xi}_c ,\ c = 1,\dots,C$, is equivalent to maximize expression \eqref{eq: M-step regression sup}, which is equivalent to independently maximize each of these $C$ expressions:

\begin{align}
\label{eq:regression max sup}
Q(\bm{\xi}_c) = \sum_{j = 1}^{J}  \tilde{\bm{z}}_{jc} \log\big{(}p(\mathbf{y}_{j}|\mathbf{x}_{j},\bm{\xi}_{c})\big{)} \quad  c = 1,...,C
\end{align}
\noindent
The maximization of \eqref{eq:regression max sup}, in each cluster, is equivalent to the maximization problem of a mixed-effects generalized linear model with a binary response and a single level of grouping induced by the $J$ groups. Only observation pairs $(\mathbf{x}_{ij},y_{ij})$ that belongs to the cluster $c$ contribute to the model.
It is important to highlight that each cluster will possess its own distinct set of parameters (fixed and random effects).

\subsection{Maximization of the Continuous covariates component - Calculation}
\label{subsec: max second term}
In each cluster, we model the continuous covariates \textbf{U} with a multivariate normal distribution with parameters $\bm{\mu}_{c}$ and $\bm{\Sigma}_c$, thus:
\begin{align}
\label{eq:phi density sup}
\phi(\mathbf{u}|\bm{\mu}_{c},\bm{\Sigma}_c)  =  \frac{1}{(2 \pi)^{\frac{p}{2}} |\bm{\Sigma}_c|^{\frac{1}{2}}} \exp\bigg{(}-\frac{1}{2} (\mathbf{u} - \bm{\mu}_{c})^T \bm{\Sigma}^{-1}_c (\mathbf{u} - \bm{\mu}_{c})\bigg{)} \quad c = 1,...,C
\end{align}
\noindent
The maximization of $Q(\bm{\theta},\bm{\theta}^{(k)})$ with respect to $\bm{\mu}_c$ and $\bm{\Sigma}_c$, $c = 1,\dots,C$, is equivalent to maximize expression \eqref{eq: M-step continuous sup}, which is equivalent to independently maximize each of these $C$ expressions (Ingrassia et al., 2015):

\begin{align}
\label{eq: cont max sup}
Q(\bm{\mu}_c;\bm{\Sigma}_c) = \sum_{j = 1}^{J} \sum_{i = 1}^{n_j} \tilde{z}_{ijc}  \log\big{(}\phi(\mathbf{u}_{ij}|\bm{\mu}_{c},\bm{\Sigma}_c)\big{)} \quad c = 1,\dots,C
\end{align}
To perform the maximization, we can rewrite \eqref{eq: cont max sup} in the following ways for $c = 1,\dots,C$:
\begin{align}
\label{rearranged phi sup}
Q(\bm{\mu}_c;\bm{\Sigma}_c)  =\sum_{j = 1}^{J} \sum_{i = 1}^{n_j} \tilde{z}_{ijc} \bigg{[}-\frac{p}{2}\log(2 \pi) -\frac{1}{2}\log(|\bm{\Sigma}_c|) -\frac{1}{2} (\mathbf{u}_{ij} - \bm{\mu}_{c})^T \bm{\Sigma}^{-1}_c (\mathbf{u}_{ij} - \bm{\mu}_{c}) \bigg{]}
\end{align}
\begin{align}
\label{rearranged phi 2 sup}
Q(\bm{\mu}_c,\bm{\Sigma}_c) =\sum_{j = 1}^{J} \sum_{i = 1}^{n_j} \tilde{z}_{ijc} \bigg{[}-\frac{p}{2}\log(2 \pi) +\frac{1}{2}\log(|\bm{\Sigma}_c^{-1}|) -\frac{1}{2} \Tr\bigg\{(\mathbf{u}_{ij} - \bm{\mu}_{c})^T  (\mathbf{u}_{ij} - \bm{\mu}_{c}) \bm{\Sigma}^{-1}_c \bigg\}  \bigg{]} 
\end{align}

\noindent
Derivative of \eqref{rearranged phi sup} with respect to $\bm{\mu}_c$:
\begin{align}
\label{eq: derivative mu sup}
 \frac{\partial Q}{\partial \bm{\mu}_{c}} = \sum_{j = 1}^{J} \sum_{i = 1}^{n_j} \tilde{z}_{ijc}  \Sigma^{-1}_c (\mathbf{u}_{ij} - \bm{\mu}_{c}) = 0 \iff \sum_{j = 1}^{J} \sum_{i = 1}^{n_j} \tilde{z}_{ijc} \mathbf{u}_{ij} = \sum_{j = 1}^{J} \sum_{i = 1}^{n_j} \tilde{z}_{ijc} \bm{\mu}_{c} 
\end{align}

\noindent
Derivative of \eqref{rearranged phi 2 sup} with respect to  $\bm{\Sigma}_c$:
 \begin{align}
 \frac{\partial Q}{\partial \bm{\Sigma}_c^{-1}} &=
 \sum_{j = 1}^{J} \sum_{i = 1}^{n_j} \tilde{z}_{ijc} \bigg{[} \frac{1}{2}\ \bm{\Sigma}_c -\frac{1}{2} (\mathbf{u}_{ij} - \bm{\mu}_{c})  (\mathbf{u}_{ij} - \bm{\mu}_{c})^T \bigg{]} = 0 \iff \notag \\
& \iff \sum_{j = 1}^{J} \sum_{i = 1}^{n_j} \tilde{z}_{ijc} \bm{\Sigma}_c = \sum_{j = 1}^{J} \sum_{i = 1}^{n_j} \tilde{z}_{ijc} (\mathbf{u}_{ij} - \bm{\mu}_{c})  (\mathbf{u}_{ij} - \bm{\mu}_{c})^T \label{eq: derivative sigma sup}
\end{align}

\noindent
From \eqref{eq: derivative mu sup} and \eqref{eq: derivative sigma sup} we obtain the update rules for $\bm{\mu}_c$ and $\bm{\Sigma}_c$:
\begin{align}
\label{a: update mu sup}
\bm{\mu}_{c}^{(k+1)} = \frac{1}{n_{c}^{(k)}} \sum_{j = 1}^{J} \sum_{i = 1}^{n_{j}} \tilde{z}_{ijc}^{(k)}\mathbf{u}_{ij} 
\end{align}

\begin{align}
\label{a: update sigma sup}
\bm{\Sigma}_{c}^{(k+1)} = \frac{1}{n_{c}^{(k)}} \sum_{j = 1}^{J} \sum_{i = 1}^{n_{j}} \tilde{z}_{ijc}^{(k)}(\mathbf{u}_{ij}-\bm{\mu}_{c}^{(k+1)})(\mathbf{u}_{ij}-\bm{\mu}_{c}^{(k+1)})^T 
\end{align}
\noindent
where, $n_c = \sum_{j = 1}^{J} \sum_{i = 1}^{n_j} \tilde{z}_{ijc}$.

\subsection{Maximization of the Categorical covariates component - Calculation}
\label{subsubsec: max third term}
The maximization of $Q(\bm{\theta},\bm{\theta}^{(k)})$ with respect to $\bm{\lambda}_c$, $c = 1,\dots,C$, is equivalent to maximize expression \eqref{eq: M-step categorical}, which is equivalent to independently maximize each of these C expressions:
\begin{align}
\label{eq:mult max sup}
\sum_{j = 1}^{J} \sum_{i = 1}^{n_j} \tilde{z}_{ijc} \log\big{(}\psi(\mathbf{v}_{ij}|\bm{\lambda}_{c})\big{)} \quad c = 1,\dots,C
\end{align}
where,
\begin{align}
\label{eq:multinom distribution sup}
\psi(\bm{v}|\bm{\lambda}_{c}) = \prod_{r = 1}^{q} \prod_{s = 1}^{k_r}  {\lambda_{crs}}^{v^{rs}} \quad c = 1,\dots,C
\end{align}
By looking at \eqref{eq:multinom distribution sup}, we can rewrite \eqref{eq:mult max sup} as:
\begin{align}
\label{eq:mult max 2 sup}
\psi(\mathbf{v}_{ij}|\bm{\lambda}_{c}) = \sum_{r = 1}^{q} \sum_{j = 1}^{J} \sum_{i = 1}^{n_j} \tilde{z}_{ijc} \sum_{s = 1}^{k_r} v^{rs}_{ij} \log{\lambda_{crs}} \quad c = 1,\dots,C
\end{align}
To simplify our notations throughout this section, we have made the assumption that each categorical covariate assumes $k_r$ categories across all the C clusters. However, in practice, our algorithm can effectively handle categorical variables with varying numbers of categories.
In each cluster $c = 1,\dots,C$ , the maximization of this function with respect to $\bm{\lambda}_{cr}$, $r = 1,\dots,q$, 
 $c = 1,\dots,C$, subject to the constraints on these parameters, is equivalent to independently maximizing each of these q expressions \citep{Ingrassia2015}:
 \begin{align}
 \label{eq: mult max 3 sup}
Q(\bm{\lambda}_{cr};\eta) = \sum_{j = 1}^{J} \sum_{i = 1}^{n_j} \tilde{z}_{ijc} \sum_{s = 1}^{k_r} v^{rs}_{ij} \log{\lambda_{crs}} - \eta \bigg{(} \sum_{s= 1}^{k_r} \lambda_{crs} - 1 \bigg{)} \quad r = 1,\dots,q
\end{align}
Derivatives:
\begin{subequations}
    \label{eq: mult deriv sup}
    \begin{align}[left=\empheqlbrace]
    \frac{\partial Q}{\partial \bm{\lambda}_{cr}}  &= \sum_{s = 1}^{k_r} \bigg{(} \sum_{j = 1}^{J} \sum_{i = 1}^{n_j} \tilde{z}_{ijc}  v^{rs}_{ij} \frac{1}{\lambda_{crs}} - \eta \bigg{)} = 0
    \label{eq: mult deriv 1 sup} \\
    \frac{\partial Q}{\partial \eta}  &= 1 - \sum_{s= 1}^{k_r} \lambda_{crs} = 0\label{eq: mult deriv 2 sup}
    \end{align}
\end{subequations}
From \eqref{eq: mult deriv 1 sup} we have:
\begin{align}
\label{eq: mult deriv 4 sup}
\lambda_{crs} = \frac{1}{\eta} \sum_{j = 1}^{J} \sum_{i = 1}^{n_j} \tilde{z}_{ijc}  v^{rs}_{ij} 
\end{align}
Substituting \eqref{eq: mult deriv 4 sup} in \eqref{eq: mult deriv 2 sup}, we obtain:
 \begin{align}
 \label{eq: mult deriv 5 sup}
\eta = \sum_{s = 1}^{k_r} \sum_{j = 1}^{J} \sum_{i = 1}^{n_j} \tilde{z}_{ijc}  v^{rs}_{ij} = \sum_{j = 1}^{J} \sum_{i = 1}^{n_j} \tilde{z}_{ijc} \bigg{(} \sum_{s = 1}^{k_r} v^{rs}_{ij} \bigg{)} = \sum_{j = 1}^{J} \sum_{i = 1}^{n_j} \tilde{z}_{ijc} = n_c
\end{align}
Replacing $\eta$ found in \eqref{eq: mult deriv 5 sup} in \eqref{eq: mult deriv 4 sup}, we obtain the update rule for $\bm{\lambda}_{cr}$:
\begin{align}
\label{a: update lambda sup}
\bm{\lambda}_{cr}^{(k+1)} = \frac{1}{n_c^{(k)}} \sum_{j = 1}^{J} \sum_{i = 1}^{n_j}   \tilde{z}_{ijc}^{(k)} v^{rs}_{ij} \quad c = 1, \dots, C, \ r = 1, \dots,q
\end{align}

\subsection{Maximization of the Dichotomous dependent covariates component}
\label{subsubsec: max fourth term}
The maximization of $Q(\bm{\theta},\bm{\theta}^{(k)})$ with respect to  $\bm{\Gamma}_c$ and $\bm{\nu}_c$, $c = 1,\dots,C$, is equivalent to maximize expression \eqref{eq: M-step dich dep sup}, which is equivalent to independently maximize each of these C expressions:
\begin{align}
Q(\bm{\Gamma}_c;\bm{\nu}_c) = \sum_{j = 1}^{J} \sum_{i = 1}^{n_j} \tilde{z}_{ijc}  \log\big{(}\zeta(\mathbf{d}_{ij}|\bm{\Gamma}_c,\bm{\nu}_c)\big{)} \quad c = 1,\dots,C
\end{align}
We describe the procedure of maximization in both domains, i.e. \{0,1\} and  \{-1,1\}. \\
The computation of the normalizing constant $Z$ involves the sum of $2^h$ terms and, for this reason, it requires a high computational cost. However, it is possible to calculate $\forall l = 1,\dots,h$ the conditional probability of $d_{ijl}$  given all the others $d_{ijk}, \ k \neq l$,  for each observation $i$ in group $j$ and in cluster $c$, without having to compute the normalizing constant (Ghosal, and Mukherjee (2020)). The key difference between the maximization procedures in the two domains lies in the exact formulation of the following formula. In the \{0,1\} domain, the conditional probability of $d_{ijl}$  given all the others $d_{ijk}, \ k \neq l$ assumes the subsequent form:
\begin{align}
\label{eq: dich 1 sup}
\mathbb{P}(d_{ijl}| d_{ijk}, k \neq l; \bm{\Gamma}, \bm{\nu} ) = \frac{\exp\big\{ d_{ijl} \cdot m_l(\bm{d}_{ij}) + \nu_{l} \cdot d_{ijl}\big\} }{\exp\big\{ m_l(\bm{d}_{ij}) + \nu_{l}\big\} + 1} 
\end{align}
Conversely, in the \{-1,1\} domain, it is expressed as follows:
\begin{align}
\label{eq: dich 1 2nd domain sup}
\mathbb{P}(d_{ijl}| d_{ijk}, k \neq l; \bm{\Gamma}, \bm{\nu} ) = \frac{\exp\big\{ d_{ijl} \cdot m_l(\bm{d}_{ij}) + \nu_{l} \cdot d_{ijl}\big\} }{\exp\big\{ m_l(\bm{d}_{ij}) + \nu_{l}\big\} + \exp\big\{ -m_l(\bm{d}_{ij}) - \nu_{l}\big\}  } 
\end{align}
where, $m_l(\bm{d}_{ij}) =  \sum_{k = 1}^{l} \gamma_{l,k} d_{ijk}$. 
Therefore, in both instances the probability of observing $\bm{d}_{ij} = ({d}_{ij1},\dots,{d}_{ijh})$ can be approximated in this way:
\begin{align}
\label{eq: dich 2 sup}
p(\bm{d}_{ij} | \bm{\Gamma}, \bm{\nu}) = \prod_{l = 1}^{h} \mathbb{P}(d_{ijl}| d_{ijk}, k \neq l )
\end{align}
Consequently, in the maximization of $Q(\bm{\theta},\bm{\theta}^{(k)})$ with respect to  $\bm{\Gamma}_c$ and $\bm{\nu}_c$, $c = 1,\dots,C$, we replace the true log-likelihood with the so-called pseudo log-likelihood:
\begin{align}
\label{eq: dich 3 sup}
Q(\bm{\Gamma}_c;\bm{\nu}_c) =  \sum_{j = 1}^{J} \sum_{i = 1}^{n_j} \tilde{z}_{ijc}  \log\big{(}p(\bm{d}_{ij}| \bm{\Gamma}_c, \bm{\nu}_c)\big{)} \quad c = 1,\dots,C 
\end{align}
Therefore, in each cluster, we found the optimal parameters $\bm{\Gamma}_c$ and $\bm{\nu}_c$ by numerically maximizing this expression.

\subsection{Maximization of the Mixture weights component}
\label{appendix: mix}
The maximization of $Q(\bm{\theta},\bm{\theta}^{(k)})$ with respect to $\bm{w}$ while satisfying the given constraints on these parameters is achieved by maximizing the augmented function \citep{Ingrassia2015}:
\begin{align}
\label{eq: max w sup}
Q(\mathbf{w};\eta) = \sum_{j = 1}^{J} \sum_{i = 1}^{n_j} \sum_{c= 1}^{C}  \tilde{z}_{ijc} \log\big{(}w_{c}\big{)} - \eta \bigg{(} \sum_{c= 1}^{C} w_c - 1 \bigg{)}
\end{align}
Derivatives:
\begin{subequations}
    \label{eq: w deriv sup}
    \begin{align}[left=\empheqlbrace]
    \frac{\partial Q}{\partial \mathbf{w}} &= \sum_{c= 1}^{C} \bigg{(}\sum_{j = 1}^{J} \sum_{i = 1}^{n_j}   \tilde{z}_{ijc} \frac{1}{w_c} - \eta \bigg{) = 0}
    \label{eq: w deriv 1 sup} \\
    \frac{\partial Q}{\partial \eta} &= \sum_{c= 1}^{C} w_c - 1 = 0
   \label{eq: w deriv 2 sup}
    \end{align}
\end{subequations}
From \eqref{eq: w deriv 1 sup}, we have:
 \begin{align}
 \label{eq: w deriv 3 sup}
w_c = \frac{1}{\eta} \sum_{j = 1}^{J} \sum_{i = 1}^{n_j} \tilde{z}_{ijc}
\end{align}
Substituting $w_c$ found in \eqref{eq: w deriv 2 sup}:
 \begin{align}
 \label{eq: w deriv 4 sup}
\eta = \sum_{j = 1}^{J} \sum_{i = 1}^{n_j} \sum_{c= 1}^{C}  \tilde{z}_{ijc} = N := \text{number of data points}
\end{align}
Replacing $\eta$ found in \eqref{eq: w deriv 4 sup} in \eqref{eq: w deriv 3 sup}, we obtain the update rule for $\bm{w}$:
\begin{align}
\label{a: update w sup}
w_c^{(k+1)} = \frac{1}{N} \sum_{j = 1}^{J} \sum_{i = 1}^{n_j}   \tilde{z}_{ijc}^{(k)} \quad c = 1, \dots, C
\end{align}

\subsection{Prediction on a new observation}
The formula for generating a prediction for a new observation with our model is derived as follows:

\begin{align}
\label{eq:joint density sup}
p((\mathbf{x},\mathbf{y})| \bm{\theta}) &= \sum_{c= 1}^{C}
p(\mathbf{y}|\mathbf{x},\bm{\xi}_{c})\phi(\mathbf{u}|
\bm{\mu}_{c},\bm{\Sigma}_c) \psi(\mathbf{v}|\bm{\lambda}_{c}) \zeta(\bm{d}| \bm{\Gamma}_c,\bm{\nu}_c)  w_{c} 
\end{align}

\begin{align}
\label{eq:pred 1 sup}
p( y | \mathbf{x}; \bm{\theta}) &= \frac{p( y; \mathbf{x}; \bm{\theta})}{p(\mathbf{x}; \bm{\theta})},
\end{align}

\noindent
From Equation~\eqref{eq:joint density sup} we have:
\begin{align}
\label{eq:pred 2 sup}
p( \mathbf{x},\mathbf{y}, \bm{\theta}) &= \sum_{c= 1}^{C}
p(\mathbf{y}|\mathbf{x},\bm{\xi}_{c})\phi(\mathbf{u}|
\bm{\mu}_{c},\bm{\Sigma}_c) \psi(\mathbf{v}|\bm{\lambda}_{c}) \zeta(\bm{d}| \bm{\Gamma}_c,\bm{\nu}_c)  w_{c} 
\end{align}
Therefore, we calculate $p(\mathbf{x}; \bm{\theta})$ as:
\begin{align}
\label{eq:pred 3 sup}
p(\mathbf{x}; \bm{\theta}) &= \int_{y} p( \mathbf{x},\mathbf{y}, \bm{\theta}) \ dy = \int_{y} \sum_{c= 1}^{C}
p(\mathbf{y}|\mathbf{x},\bm{\xi}_{c}) \phi(\mathbf{u}|
\bm{\mu}_{c},\bm{\Sigma}_c) \psi(\mathbf{v}|\bm{\lambda}_{c}) \zeta(\bm{d}| \bm{\Gamma}_c,\bm{\nu}_c)  w_{c} \ dy = \notag \\ 
&= \sum_{c= 1}^{C} \phi(\mathbf{u}|
\bm{\mu}_{c},\bm{\Sigma}_c) \psi(\mathbf{v}|\bm{\lambda}_{c}) \zeta(\bm{d}| \bm{\Gamma}_c,\bm{\nu}_c)  w_{c} \int_{y} p(\mathbf{y}|\mathbf{x},\bm{\xi}_{c}) \ dy = \notag \\
&= \sum_{c= 1}^{C} \phi(\mathbf{u}|
\bm{\mu}_{c},\bm{\Sigma}_c) \psi(\mathbf{v}|\bm{\lambda}_{c}) \zeta(\bm{d}| \bm{\Gamma}_c,\bm{\nu}_c)  w_{c}
\end{align}
At this point, to derive the prediction, we calculate the expected value of $p( y | \mathbf{x}; \bm{\theta})$ by substituting Equation~\eqref{eq:pred 2 sup} and~\eqref{eq:pred 3 sup} into~\eqref{eq:pred 1 sup}:
\begin{align}
\label{eq:pred 4 sup}
\mathbb{E}[ y | \mathbf{x}; \bm{\theta}] = p( y = 1 | \mathbf{x}; \bm{\theta}) &= \frac{\sum_{c= 1}^{C} \phi(\mathbf{u}|
\bm{\mu}_{c},\bm{\Sigma}_c) \psi(\mathbf{v}|\bm{\lambda}_{c}) \zeta(\bm{d}| \bm{\Gamma}_c,\bm{\nu}_c)  w_{c} \cdot p(y = 1|\mathbf{x},\bm{\xi}_{c})}{\sum_{c= 1}^{C} \phi(\mathbf{u}|
\bm{\mu}_{c},\bm{\Sigma}_c) \psi(\mathbf{v}|\bm{\lambda}_{c}) \zeta(\bm{d}| \bm{\Gamma}_c,\bm{\nu}_c)  w_{c}}.
\end{align}

\subsection{Pseudo-code of the Algorithm}
\label{appendix: pseudo-code algo}
\makeatletter
\renewcommand{\fnum@algorithm}{}
\renewcommand{\ALG@name}{}
\makeatother
\captionsetup[algorithm]{
justification=centering,
    font={color=black, bf} 
}
Let $\bm{\theta} = (w_1, \dots,  w_c,\ \bm{\beta}_1,\dots,\bm{\beta}_c, \ \sigma_{b1},\dots,\sigma_{bc}, \ \bm{\mu}_1,\dots,\bm{\mu}_c,\ \bm{\Sigma}_1,\dots,\bm{\Sigma}_c,\ \bm{\lambda}_1,\dots,\bm{\lambda}_c, 
\bm{\Gamma}_1,\dots, \bm{\Gamma}_c, \bm{\nu}_1,\dots,\bm{\nu}_c)$ be the vector containing all the parameters of the model. Note that $\bm{\xi}_c = (\bm{\beta_c}, \sigma_{bc})$.
\begin{algorithm}[H]
\caption{E-step}
\begin{algorithmic}[1]
\State \textbf{Input} $\rightarrow$ Data, $\bm{\theta}$, C
\For{c in 1:C}
     \State Compute $\tau_{ijc}$ with the formula: \begin{align}
\mathbb{E}\big{[} z_{ijc}|(\mathbf{x},\mathbf{y}), \hat{\bm{\theta}}^{(k)} \big{]} = \mathbb{P}\big{[} z_{ijc} = 1|(\mathbf{x}_{ij},y_{ij}, ), \hat{\bm{\theta}}^{(k)} \big{]} =  \tau_{ijc}^{(k)}((\mathbf{x}_{ij},y_{ij}), \hat{\bm{\theta}}^{(k)}) = \notag \\
= \frac{p(y_{ij}|\mathbf{x}_{ij},\hat{\bm{\xi}}_{c}^{(k)})\phi(\mathbf{u}_{ij}|\hat{\bm{\mu}}_{c}^{(k)},\hat{\bm{\Sigma}}_c^{(k)}) \psi(\mathbf{v}_{ij}|\hat{\bm{\lambda}}_{c}^{(k)}) \zeta(\bm{d}_{ij}| \hat{\bm{\Gamma}}_c^{(k)},\hat{\bm{\nu}}_c^{(k)}) \hat{w}_{c}^{(k)}} {\sum_{c = 1}^{C} p(y_{ij}|\mathbf{x}_{ij},\hat{\bm{\xi}}_{c}^{(k)})\phi(\mathbf{u}_{ij}|\hat{\bm{\mu}}_{c}^{(k)},\hat{\bm{\Sigma}}_c^{(k)}) \psi(\mathbf{v}_{ij}|\hat{\bm{\lambda}}_{c}^{(k)}) \zeta(\bm{d}_{ij}| \hat{\bm{\Gamma}}_c^{(k)},\hat{\bm{\nu}}_c^{(k)}) \hat{w}_{c}^{(k)}}, \label{eq:E-step sup}
\end{align}
\EndFor
\State Compute \textbf{z} by assigning each observation to the cluster with the maximum probability
\State \Return \textbf{z}
\end{algorithmic}
\end{algorithm}
\noindent
NOTE: $\tau_{ijc}$ is the probability of each observation of belonging to the latent group c
\begin{algorithm}[H]
\caption{M-step}
\begin{algorithmic}[1]
\State \textbf{Input} $\rightarrow$  Data, \textbf{z}, C 
\State {\ttfamily\textcolor{customblue}{/* Update the parameters related to Y */} }
\For{c in 1:C}
      \State {Update $\bm{\xi}_c$ using command $\texttt{lme4::glmer}^1$ with weights $z_{ijc}$}
\EndFor
\State {\ttfamily\textcolor{customblue}{/* Update the parameters related to U */} }
\For{c in 1:C}
    \State {Compute $\bm{\mu}_c$ and $\bm{\Sigma}_c$ using formulas \eqref{a: update mu sup} and \eqref{a: update sigma sup}, respectively} 
\EndFor
\State {\ttfamily\textcolor{customblue}{/* Update the parameters related to V */} }
\For{c in 1:C}
    \State {Compute $\bm{\lambda}_c$ using formula \eqref{a: update lambda sup}} 
\EndFor
\State {\ttfamily\textcolor{customblue}{/* Update the parameters related to D */} }
\For{c in 1:C}
    \State {Update $\bm{\Gamma}_c$ and $\bm{\nu}_c$ with command $\texttt{EstimateIsing}^2$ or $\texttt{IsingFit}^3$ using only observations for which $\tilde{z}_{ijc} = 1$} 
\EndFor
\State {\ttfamily\textcolor{customblue}{/* Update mixture weights */} }
\For{c in 1:C}
\State {Compute $w_c$ using formula \eqref{a: update w sup}} 
\EndFor
\State \Return $\bm{\theta} $
\end{algorithmic}
\end{algorithm}
\noindent
\begin{enumerate}
    \item \texttt{lme4::glmer(formula=formula,data=data,family="binomial",weights=z[,c])}
    \item \texttt{IsingSampler::EstimateIsing(D[which(z[,c] == 1),],beta = 1,responses = c(0L, 1L), method = "pl")}:this command estimates the optimal parameters of the Ising model by maximizing the pseudo-likelihood \eqref{eq: dich 3 sup}
    \item  \texttt{IsingFit::IsingFit(D[\text{which}(z[,c] == 1),], family="binomial")}: this command perform the network estimation of the Ising model by using the eLasso method. It  combines l1-regularized logistic regression with model selection based on the EBIC.  
\end{enumerate}

\begin{algorithm}[H]
\caption{EM-Algorithm for ML-CWMd}
\begin{algorithmic}[1]
\State Initialize $\bm{\theta}$ = $\bm{\theta}^{(0)}$ with k-means
\State Select the number of assumed latent clusters C
\State Select a tolerance, for example tol = $10^{-5}$
\State k = 1
\While{$l((\mathbf{x},\mathbf{y}, \mathbf{z}); \bm{\theta}^{(k)})$ - $l((\mathbf{x},\mathbf{y}, \mathbf{z}); \bm{\theta}^{(k-1)})$ $<$ tol}
    \State \textbf{z} = \textbf{E-step}(Data, $\bm{\theta}^{(k-1)}$, C) 
    \State $\bm{\theta}^{(k)}$ = \textbf{M-step}(Data; \textbf{z}, C) 
    \State Compute $l((\mathbf{x},\mathbf{y}, \mathbf{z}); \bm{\theta}^{(k)})$ 
     \State k = k + 1
\EndWhile   
\State \Return \textbf{z} and \ $\bm{\theta}^{(k)}$
\end{algorithmic}
\end{algorithm}

\newpage

\section*{Web Appendix C: Further results pertaining the simulation study}
\begin{figure}[H]
    \centering
    \begin{subfigure}[b]{0.49\textwidth}
        \centerline{
        \includegraphics[width=\textwidth]{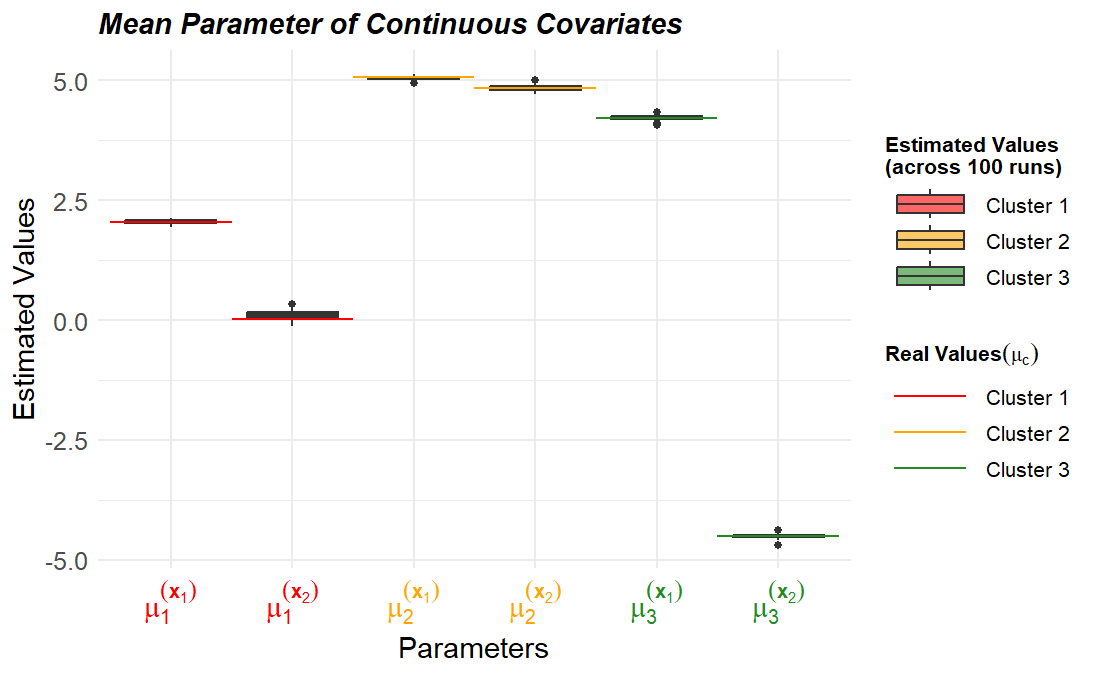}}
        \caption{Comparison of the distributions of estimated mean parameters $\bm{\mu}_c$ ($c = 1, \ \hdots, C$) across the simulation experiment with the true values related to continuous covariates ($\textbf{x}_1$, $\textbf{x}_2$).}
        \label{fig:mu-par}
    \end{subfigure}
    \hfill
    \begin{subfigure}[b]{0.49\textwidth}
        \centerline{
        \includegraphics[width=\textwidth]{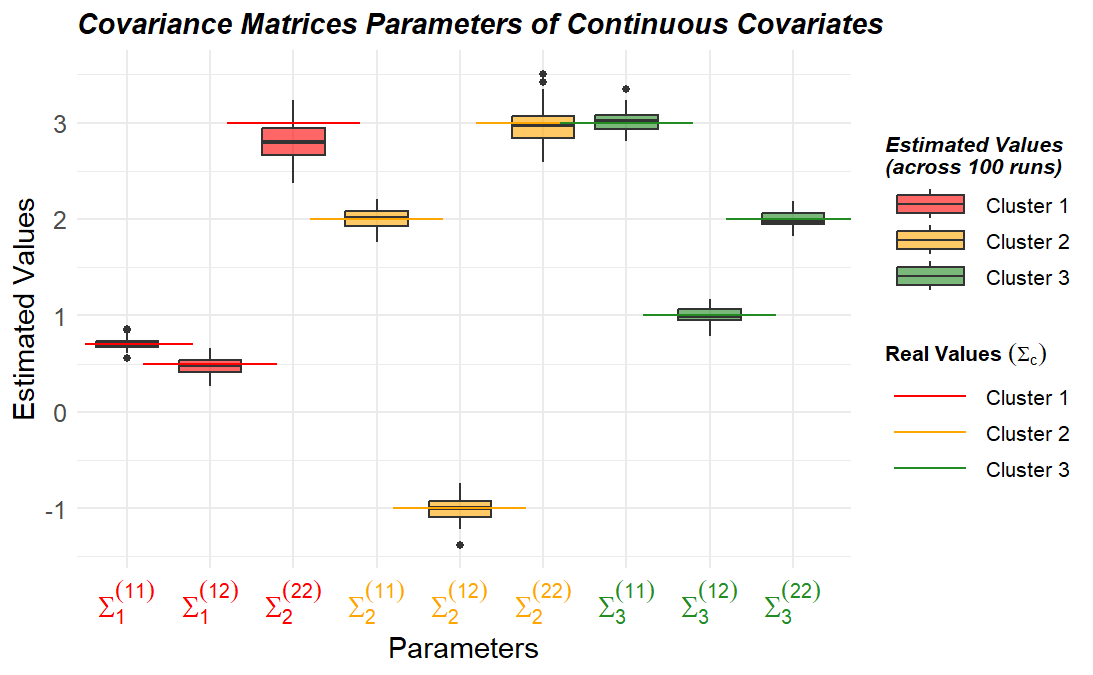}} 
        \caption{Comparison of the distributions of estimated covariance matrices parameters $\bm{\Sigma}_c$ ($c = 1, \ \hdots, C$) across the simulation experiment with the true values related to continuous covariates ($\textbf{x}_1$, $\textbf{x}_2)$.}
        \label{fig:cov-par}
    \end{subfigure}
    \begin{subfigure}[b]{0.49\textwidth}
        \centerline{
        \includegraphics[width=\textwidth]{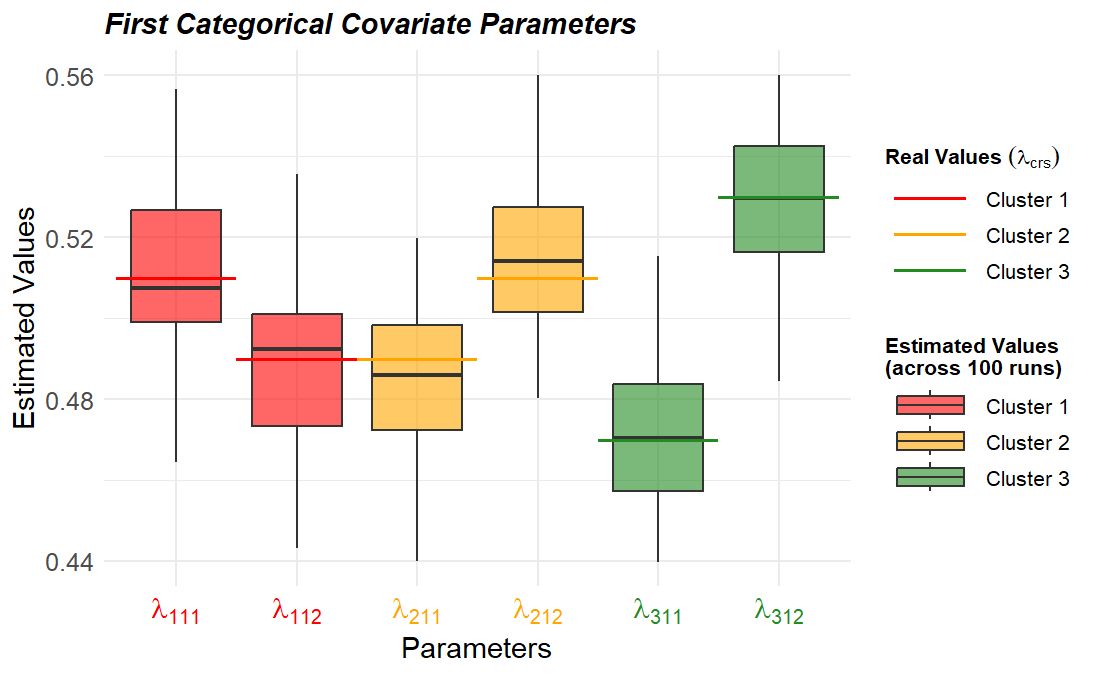}}
        \caption{Comparison of the distributions of estimated parameters $\bm{\lambda}_{crs}$ ($c = 1, \ \hdots, C$, $r = 1, \ \hdots, q$, $s = 1, \ \hdots, k_r$) across the simulation experiment with the true values related to the first categorical covariate $\textbf{a}_1$.}
        \label{fig:lam1-par}
    \end{subfigure}
    \hfill
    \begin{subfigure}[b]{0.49\textwidth}
        \centerline{
        \includegraphics[width=\textwidth]{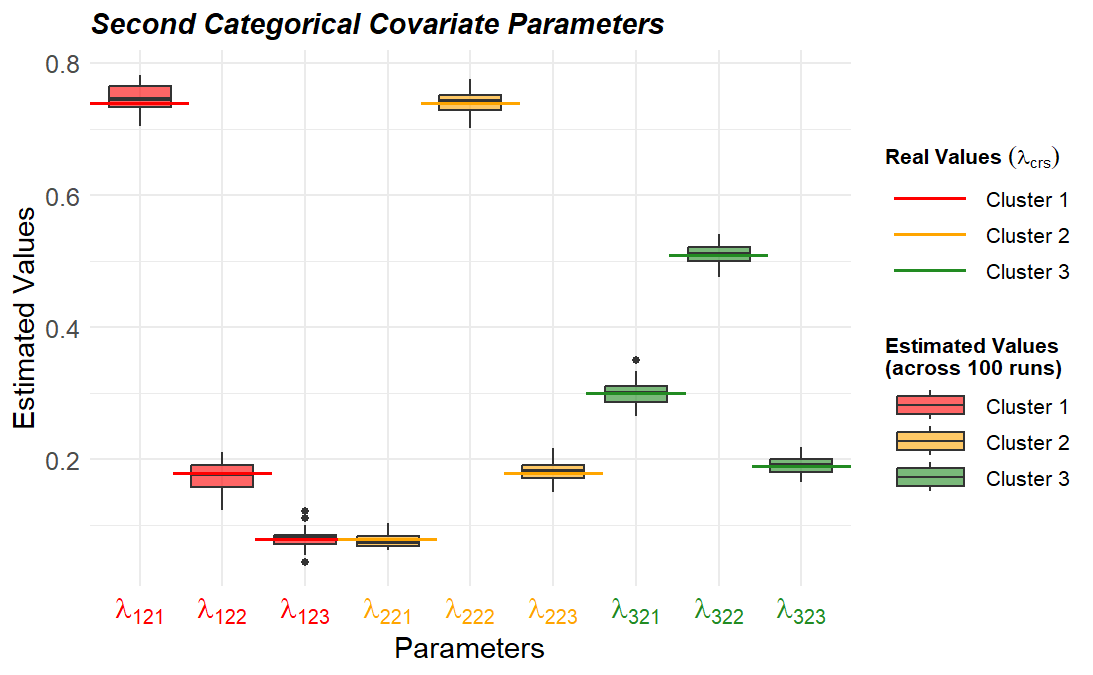}} 
        \caption{Comparison of the distributions of estimated parameters $\bm{\lambda}_{crs}$ ($c = 1, \ \hdots, C$, $r = 1, \ \hdots, q$, $s = 1, \ \hdots, k_r$) across the simulation experiment with the true values related to the first categorical covariate $\textbf{a}_2$.}
        \label{fig:lam2-par}
    \end{subfigure}
    \caption{}
    \label{par_rec1}
\end{figure}

\begin{figure}[htbp]
    \centering
    \begin{subfigure}[b]{0.45\textwidth}
        \centerline{
        \includegraphics[width=\textwidth]{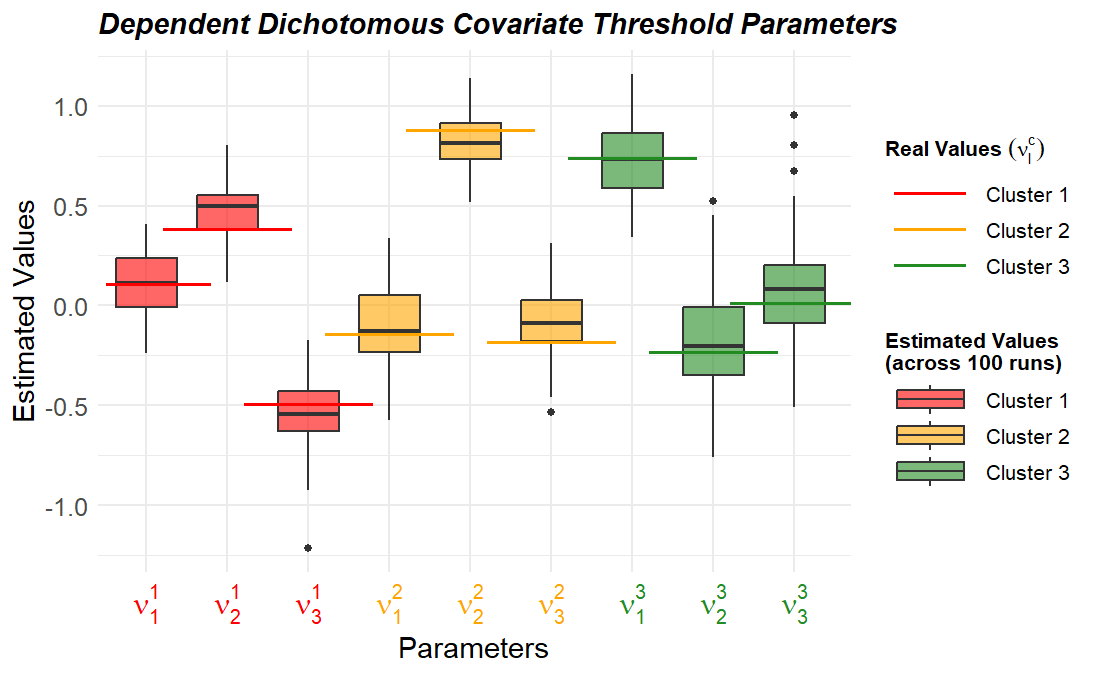}} 
        \caption{Comparison of the distributions of estimated threshold parameter vectors $\bm{\nu}_{l}^c$ ($l = 1, \ \hdots, h$, $c = 1, \ \hdots, C$) across the simulation experiment with the true values related to the dichotomous dependent covariates ($\textbf{d}_1$, $\textbf{d}_2$, $\textbf{d}_3$).}
        \label{fig:thr-par}
    \end{subfigure}
    \hfill
    \begin{subfigure}[b]{0.49\textwidth}
        \centerline{
        \includegraphics[width=\textwidth]{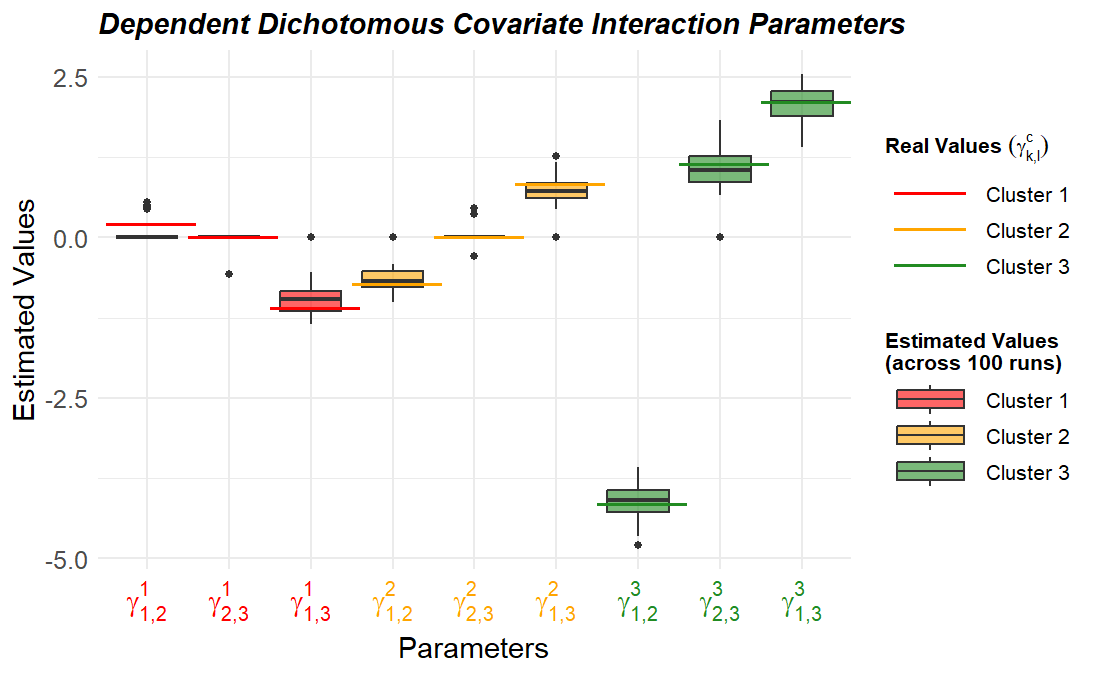}} 
        \caption{Comparison of the distributions of estimated interaction matrices parameters $\bm{\gamma}_{lk}^c$ ($l,k = 1, \ \hdots, h$, $c = 1, \ \hdots, C$) across the simulation experiment with the true values related to the dichotomous dependent covariates ($\textbf{d}_1$, $\textbf{d}_2$, $\textbf{d}_3$).}
        \label{fig:int-par}
    \end{subfigure}
    \hfill
    \begin{subfigure}[b]{0.49\textwidth}
        \centerline{
        \includegraphics[width=\textwidth]{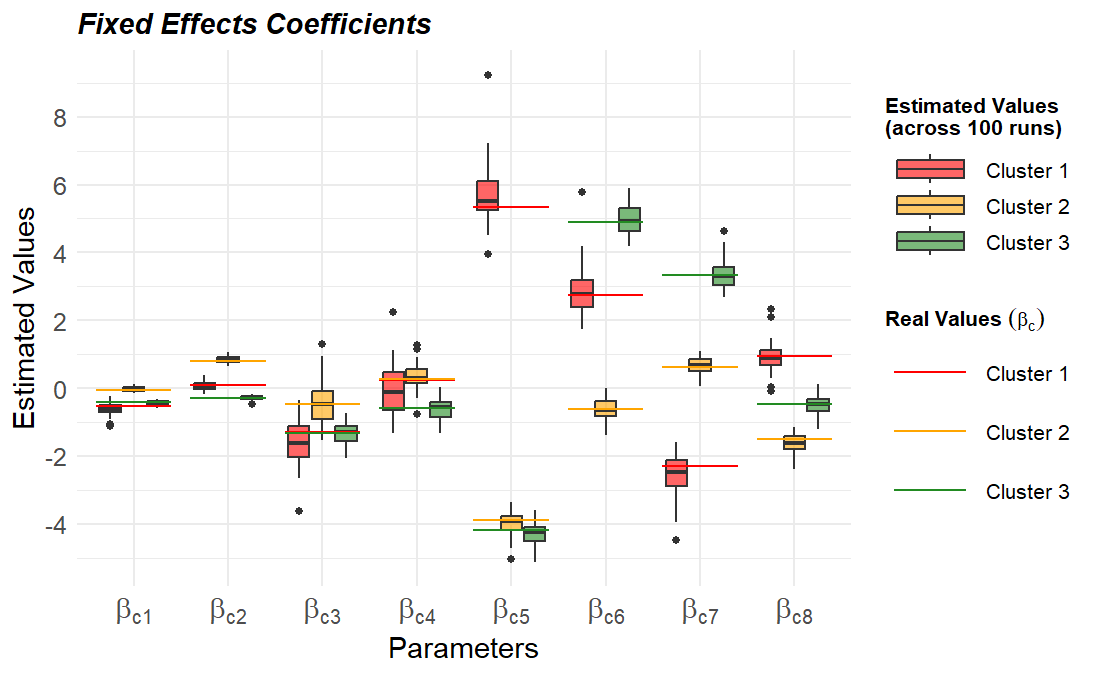}} 
        \caption{Comparison of the distributions of estimated fixed effects coefficients $\bm{\beta}_{c}$ ($c = 1, \ \hdots, C$) across the simulation experiment with the true values.}
        \label{fig:bet-par}
    \end{subfigure}
     \hfill
    \begin{subfigure}[b]{0.49\textwidth}
        \centerline{
        \includegraphics[width=\textwidth]{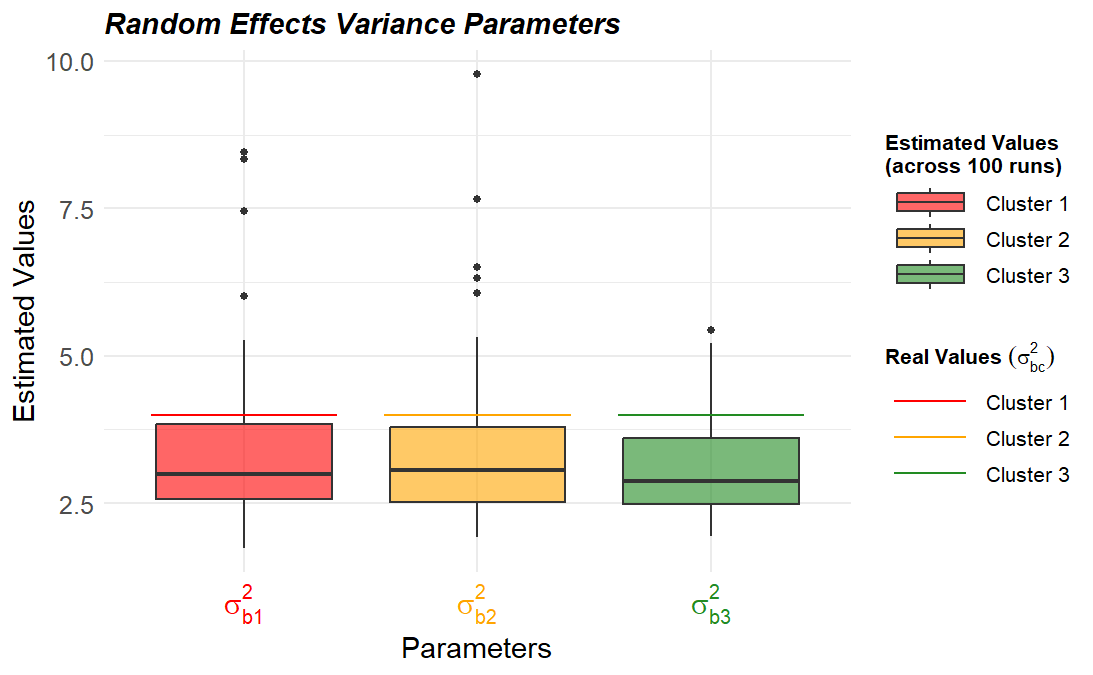}} 
        \caption{Comparison of the distributions of estimated variance of random effects $\bm{\sigma}_{bc}^2$ ($c = 1, \ \hdots, C$) across the simulation experiment with the true values.}
        \label{fig:var-par}
    \end{subfigure}
    \caption{}
    \label{par_rec2}
\end{figure}

In this appendix, we present a comparison, pertaining to the simulation study presented in the main text, between the estimated parameters' distributions from the simulation experiment and the true values associated with all covariates in the DPG. These results pertain to the ML-CWMd method, which has been identified in the paper as the most efficient approach for capturing the heterogeneity present in the DPG. In this appendix, we exclusively consider the coefficients associated with ML-CWMd for the fixed effects parameters $\bm{\beta}_c$ ($c = 1, \ \hdots, C$). This analysis mirrors the comparison presented in the main text, focusing solely on ML-CWMd coefficients. As illustrated in Figures~\ref{par_rec1} and~\ref{par_rec2}, the algorithm demonstrates proficient recovery of the true parameter values within the DPG. 
\newpage

\section*{Web Appendix D: Description of clinical variables involved in the analysis}
In the following, we provide a brief description of the respiratory diseases considered in the analysis and of the Multi-source Comorbidity Score (MCS). 

\noindent
The Chronic obstructive pulmonary disease (COPD) entails a persistent constriction or blockage of the air passages resulting in an ongoing reduction in the airflow rate during exhalation. Pneumonia (PNA) refers to the abrupt inflammation of the lungs brought about by an infection. Respiratory failure (RF) manifests when the oxygen level in the blood becomes critically low or when there is a dangerous elevation in the blood's carbon dioxide level. Bronchitis (BRH) entails an inflammation of the primary air passages of the lungs, known as the bronchi, typically triggered by infection, leading to irritation and inflammation.

\noindent
For what concerns the computation of the MCS, the scores associated to single diseases are presented in Table~\ref{table MCS 1}. The overall score for each patient is computed by summing the scores associated with all the diseases they have been diagnosed with. The diseases attributed to each patient are identified by analyzing their medical history pertaining the previous 5 years, a process facilitated through the utilization of ICD-9-CM codes. These codes serve as a means to extract relevant medical information and establish connections to specific diseases. 
\subsection{ICD-9-CM codes for respiratory diseases}
\begin{itemize}
    \item \textbf{Chronic obstructive pulmonary disease (COPD)}:  \texttt{491-496, 491.2, 492.0, 492.8, 494.0-494.1}
    \item \textbf{Pneumonia (PNA):} \texttt{480-486, 507, 011.6, 052.1, 055.1, 073.0, 130.4, 480.0-480.3, \\ 480.8-480.9, 482.1-482.4, 482.8-482.9, 483.1, 483.8, 484.3, 484.5, 487.0, \\ 506.0, 507.0, 507.8, 517.1, 770.0, 00322, 011.61-011.66, 115.05, 115.15, 115.95, 482.30-482.32, 482.40-482.41, 482.49, 482.81, 482.89, V12.61}
    \item \textbf{Respiratory failure (RF):} \texttt{518.81, 518.83-518.84}
    \item \textbf{Bronchitis (BRH):} \texttt{466, 490-491,  466.0, 491.0-491.2, 491.8-491.9, 491.20-491.22}
\end{itemize}

\begin{table}[htpb]
\renewcommand{\arraystretch}{1.3}
\caption{The score associated to each disease to define the Modified Multisource-Comorbidity Score (MCS).}
\begin{minipage}{0.5\textwidth}
\centering
\begin{tabular}{ l c}
\hline
\textbf{Comorbidity} & \textbf{Score} \\
\hline 
Metastatic cancer & 18  \\
\hline
Alcohol abuse & 11 \\
\hline
Non-metastatic cancer & 10 \\
  \hline
    Tuberculosis & 10    \\
    \hline
    Psychosis & 8     \\
    \hline
    Liver diseases & 8  \\
    \hline
    Drugs for anxiety & 6    \\
    \hline
    Weight loss & 6    \\
    \hline
    Dementia & 6    \\
    \hline
    Drugs for malignancies & 5    \\
    \hline
    Parkinson's disease & 5   \\
    \hline
    Lymphoma & 5   \\
    \hline
    Paralysis & 5    \\
    \hline
    Coagulopathy & 5   \\
    \hline
    Fluid disorders & 4    \\
    \hline
    Kidney diseases & 4    \\
    \hline
\end{tabular}
\label{table MCS 1}
\end{minipage}%
\begin{minipage}{0.5\textwidth}
\centering
\begin{tabular}{ l c}
\hline
\textbf{Comorbidity} & \textbf{Score}  \\
\hline 
 Kidney dialysis & 4    \\
    \hline
    Heart failure & 4    \\
    \hline
    Other neurological disorders & 3    \\
    \hline
    Rheumatic diseases & 3    \\
    \hline
    Brain diseases & 3   \\
    \hline
     Anemia & 3    \\
    \hline
    Diabetes & 2    \\
    \hline
    Gout & 2    \\
    \hline
    Epilepsy & 2    \\
    \hline
    Ulcer diseases & 2    \\
    \hline
Myocardial infarction & 1 \\
\hline
Drugs for coronary & 1 \\
\hline
Valvular diseases & 1 \\
\hline
Arrhythmia & 1 \\
\hline
Obesity & 1 \\
\hline
Hypothyroidism & 1 \\
\hline
\end{tabular}
\label{table MCS 2}
\end{minipage}
\end{table}






\end{document}